\documentclass[12pt]{iopart}
\usepackage{iopams}  
\usepackage{graphicx}
\usepackage{setstack}
\usepackage{color}

\renewcommand{\mat}[1]{\mathbb{#1}}
\renewcommand{\vec}[1]{\bi{#1}}
\newcommand{\vk}{\vec{k}}

\newcommand{\vp}{\vec{p}}
\newcommand{\eqref}[1]{(\ref{#1})}

\newcommand{\G}{\mathbb{G}}

\begin{document}

\title[Superconductivity in the attractive
  Hubbard model: fRG]{Superconductivity in the attractive
  Hubbard model: functional renormalization group analysis}

\author{R Gersch$^1$, C Honerkamp$^2$ and W Metzner$^1$}

\address{$^1$ Max-Planck-Institut f\"{u}r Festk\"{o}rperforschung, 
  Heisenbergstra\ss{}e 1, 70569 Stuttgart}
\address{$^2$ Institut f\"{u}r
  Theoretische Physik, Universit\"{a}t W\"{u}rzburg, Am Hubland, 
  D-97074 W\"{u}rzburg, Germany}

\ead{r@roland-gersch.de}
\begin{abstract}
We present a functional renormalization group analysis of
superconductivity in the ground state of the attractive Hubbard
model on a square lattice.
Spontaneous symmetry breaking is treated in a purely fermionic
setting via anomalous propagators and anomalous effective 
interactions.
In addition to the anomalous interactions arising already in the
reduced BCS model, effective interactions with three incoming 
legs and one outgoing leg (and vice versa) occur.
We accomplish their integration into the usual diagrammatic formalism
by introducing a Nambu matrix for the effective interactions.
From a random-phase approximation generalized through use of this 
matrix we conclude that the impact of the $3+1$ effective interactions 
is limited, especially considering the effective interactions 
important for the determination of the order parameter.
The exact hierarchy of flow equations for one-particle irreducible
vertex functions is truncated on the two-particle level, with
higher-order self-energy corrections included in a scheme proposed
by Katanin.
Using a parametrization of effective interactions by patches 
in momentum space, the flow equations can be integrated
numerically to the lowest scales without encountering divergences.
Momentum-shell as well as interaction-flow cutoff functions are 
used, including a small external field or a large external field 
and a counterterm, respectively.
Both approaches produce momentum-resolved order parameter values 
directly from the microscopic model. The size of the superconducting
gap is in reasonable agreement with expectations from other studies.

%%% Local Variables: 
%%% mode: latex
%%% TeX-master: "main"
%%% End: 

\end{abstract}

\maketitle

%\tableofcontents{}

\section{Introduction}
Renormalization group (RG) techniques facilitate the systematic 
investigation of interacting many-particle systems even
in situations where perturbation theory breaks down.
A central idea underlying many renormalization group
methods, pioneered by Wilson \cite{PhysRevB.4.3184}, is to
take into account the degrees of freedom successively,
ordered by their kinetic energy.
Formally, this is accomplished by integrating a set of
differential equations called flow equations, where some
energy scale (usually a cutoff) plays the role of the flow
parameter.
The RG flow generates a family of effective actions which
interpolates smoothly between the bare action of the system 
and the final effective action where all degrees of freedom
have been integrated out.

The first significant applications of RG methods to interacting
Fermi systems in more than one dimension were undertaken by
mathematicians aiming at a rigorous control of certain systems
and their properties \cite{PhysRevB.42.9967,ft1,ft2}.
The usefulness of RG concepts for Fermi systems was soon recognized 
also by physicists
\cite{1991PhyA..177..530S,1994RvMP...66..129S}.
RG calculations for Fermi systems are often termed ''functional''
as they involve the flow of functions instead of just a few 
running couplings.
An early computational success of the functional RG (fRG) was
unambiguous evidence for d-wave superconductivity in the 
repulsive two-dimensional Hubbard model
\cite{PhysRevB.61.7364,2001PhRvB..63c5109H,0295-5075-44-2-235}.

Of particular interest in interacting Fermi systems is the 
spontaneous breaking of symmetries.
Formally, such a symmetry breaking occurs if a product of 
two fermionic fields -- a bilinear -- which does not appear in
the original Hamiltonian acquires a nonzero expectation value.
This bilinear can be replaced by a bosonic field through a 
Hubbard-Stratonovich transformation, which simplifies the 
treatment of the order parameter and its fluctuations \cite{Popov}.
However, these simplifications come at the price of introducing a 
bias towards breaking the symmetry in the channel fixed by the 
Hubbard-Stratonovich field.
Working with the original fermionic degrees of freedom this bias
can be avoided. 
However, the symmetry-broken phase is not easily accessible
by a purely fermionic RG flow. Calculations starting from the
symmetric phase run into divergences at a finite energy scale,
that is before all degrees of freedom have been integrated out.

Most previous functional RG studies of interacting Fermi systems 
are based on a flow equation for the generator of the one-particle
irreducible (1PI) vertex functions \cite{2001PThPh.105....1S}.
This equation is equivalent to an infinite hierarchy of flow 
equations for the 1PI vertex functions.
Truncating the hierarchy in a weak coupling expansion, the RG
flow has been computed for various low-dimensional Fermi systems
\cite{2001PhRvB..63c5109H,andergassen:153308,andergassen:075102,
enss:155401,2006EL.....75..146F,rohe:115116}.
Instead of the common choice of a momentum cutoff, 
it was found to be advantageous in 
certain situations to use as flow parameter the physical 
temperature \cite{PhysRevB.64.184516}, the strength of the 
interaction \cite{2004PhRvB..70w5115H}, or a cutoff in frequency 
space \cite{sabinediss,tilman_diss}.

Spontaneous symmetry breaking within the functional RG framework
has frequently been studied by introducing bosonic fields for the
order parameter, leading to a coupled flow of bosons and fermions
\cite{baier:2004,schuetz:2006,philipppaper}. 
A simple way to access spontaneous symmetry breaking in a purely 
fermionic setting is to combine the fRG with a mean-field 
calculation for the low-energy degrees of freedom \cite{reiss:075110}.
However, order parameter fluctuations are thereby discarded.
To continue the fermionic functional RG flow into the symmetry-broken
phase, an improved truncation of the hierarchy with an efficient
treatment of self-energy corrections, first formulated by Katanin 
\cite{2004PhRvB..70k5109K}, is very promising.
Employing this improvement, the Bardeen-Cooper-Schrieffer (BCS) 
mean-field model can be solved exactly already by a low-order
truncation \cite{2004PThPh.112..943S}, if a small external 
field is used to trigger the symmetry-breaking.
The vaunted divergence is then regularized, but there is still
a flow regime with strong coupling.
The method was extended to the breaking of a discrete symmetry
and finite temperatures \cite{2005EPJB...48..349G}.
Employing the interaction flow \cite{2004PhRvB..70w5115H},
the external field can be compensated by a counterterm,
thus enabling the study of first-order phase transitions
which were previously out of the method's reach 
\cite{2006NJPh....8..320G}.
Furthermore, this approach yields results at zero external
field without having to pass through a strong coupling regime
in the case of discrete-symmetry breaking.
Strong attractors for the flows of various counterterm 
strengths were found.

In this work, we apply the fermionic fRG to the attractive
Hubbard model, using both the external-field as well as the 
counterterm approach.
The repulsive Hubbard model
\cite{PhysRev.137.A1726,hubbard,1963PThPh..30..275K}
was originally introduced to study ferromagnetism of 
itinerant electrons, but has since been employed to elucidate 
many aspects of strong correlations in solid-state physics.
The attractive version exhibits s-wave superconductivity 
under certain conditions, including the case we study here, 
namely for a square lattice at zero temperature and weak or 
moderate coupling \cite{RevModPhys.62.113}.
It is a good testbed because the correct order parameter 
is affected by fluctuations even in the weak coupling limit
\cite{PhysRevB.45.13008},
while the model is fairly well-studied, allowing comparisons 
of the results obtained using the new method.
Lattice fermions with attractive interactions can nowadays
be realized by cold atoms in optical lattices, and s-wave
superfluidity in such systems has been observed very recently
\cite{Chin:2006}.

Our goals in this work are threefold.
First, we would like to study the role of anomalous effective
interactions with three ingoing particles and one outgoing 
particle (or vice versa), which arise in systems with both 
Cooper and forward scattering.
We find that these interactions have a limited impact.
Second, we would like to determine whether the application of the
Katanin-truncated fRG to the Hubbard model works as well as for 
mean-field models.
For the momentum-shell approach with a small external field, 
we obtain reasonable flows resulting in order parameter
values comparable to the literature.
It is possible to choose the strength of the external field 
two orders of magnitude below the final order parameter value
without running into artificial divergences.
For the interaction-flow approach, we find remnants of the
strong-attractor behavior known from the mean-field models.
Third, we compute the superconducting order parameter in the
ground state of the attractive Hubbard model as a function 
of coupling and filling.

This paper is structured as follows.
In section \ref{sec:attractive}, the attractive Hubbard model
is introduced formally.
The formalism permitting the study of symmetry breaking
in general and the $3+1$ anomalous effective interactions in particular
is detailed in section \ref{sec:nambu}.
It is employed in the context of a generalized random-phase approximation (RPA)
of the model's behavior in section \ref{sec:mf_resum}.
Section \ref{sec:frg} presents the result of the 
fRG studies, containing momentum-shell and interaction
flows as well as a study of the order parameter strength
and a comparison to earlier calculations.

\section{Attractive Hubbard model}
\label{sec:attractive}
We discuss the attractive Hubbard model
on a two-dimensional square lattice specified by the Hamiltonian
\begin{equation}
  \label{eq:hubbard_ham}
  H=\sum_{\vec{k}s}
  \xi^{\phantom\dagger}_{\vec{k}}c_{\vec{k}s}^\dagger c^{\phantom\dagger}_{\vec{k}s}
  -
  \frac{U_0}{N}\sum_{\vec{k}_1\vec{k}_2\vec{k}_3}
  c_{\vec{k}_1\uparrow}^\dagger c^{\phantom\dagger}_{\vec{k}_3\uparrow}
  c_{\vec{k}_2\downarrow}^\dagger c^{\phantom\dagger}_{\vec{k}_1+\vec{k}_2-\vec{k}_3\downarrow}.
\end{equation}
$N$ denotes the number of lattice points, 
$U_0>0$ is the onsite attraction, 
and $\vec{k}$ takes all values in the first Brillouin zone of 
the square lattice. 
The dispersion relation
\begin{equation}
  \label{eq:hubbarddispersion}
  \xi_\vk=-2t(\cos k_x+\cos k_y)-4t^\prime\cos k_x\cos k_y-\mu
\end{equation}
contains a probability amplitude $t$ for nearest-neighbor hopping,
a probability amplitude $t^\prime$ for next-nearest-neighbor hopping,
and a chemical potential $\mu$.
An external field
\begin{equation}
  \label{eq:ext_pairing}
  \sum_{\vec{k}} \left(
  \Delta^{\phantom\star}_{\rm ext}(\vec{k}) 
  c_{-\vec{k}\downarrow}^\dagger c^{\dagger}_{\vec{k}\uparrow}
  +
  \Delta^{\star}_{\rm ext}(\vec{k})
  c^{\phantom\dagger}_{\vec{k}\uparrow} c_{-\vec{k}\downarrow}^{\phantom\dagger} 
  \right)
\end{equation}
is added to $H$ to trigger the instability towards superconductivity.
This field is either kept small 
or turned off gradually in the fRG flow
to approximate the situation with spontaneous symmetry breaking.

\section{Nambu formalism}
\label{sec:nambu}
\subsection{Functional integral and field substitution}
We study the partition function ${\cal Z}$
in a functional integral representation 
\begin{equation}
  \label{eq:partitionfunction}
  {\cal Z} = \int{\cal D}(\bar{\psi},\psi)
  e^{S(\bar{\psi},\psi)},
\end{equation}
within the Matsubara formalism,
employing the grand canonical action
\begin{eqnarray}
  \nonumber
  S=&\sum_{ks}(\rmi\omega_n-\xi_{\vec{k}})
  \bar{\psi}_{ks}\psi_{ks} 
  - \sum_k
  \left(
    \Delta_{\rm ext}(\vk)
    \bar{\psi}_{-k\downarrow}\bar{\psi}_{k\uparrow}
    +
    \Delta_{\rm ext}^*(\vk)
    \psi_{k\uparrow}\psi_{-k\downarrow}
  \right)
  \\&  +U_0\sum_{k_1,k_2,k_3}
    \bar{\psi}_{k_1\uparrow}\psi_{k_3\uparrow}
  \bar{\psi}_{k_2\downarrow}\psi_{k_1+k_2-k_3\downarrow}
  \label{eq:actionpsi}
\end{eqnarray}
where $\psi$ and $\bar{\psi}$ are the Grassmann fields matching $c$ and
$c^\dagger$, respectively.
Furthermore, $k=(\rmi\omega_n,\vec{k})$ and volume as well as 
temperature factors are implicitly contained in the summation
symbol $\sum$, a convention we also adopt in the following.
We employ the substitution 
\begin{equation}
  \label{eq:actsub}
  \bar{\phi}_{k+}=\bar{\psi}_{k\uparrow},\;
  {\phi}_{k+}=\psi_{k\uparrow},\;
  {\phi}_{k-}=\bar{\psi}_{-k\downarrow},\;
  \bar{\phi}_{k-}={\psi}_{-k\downarrow}
\end{equation}
to obtain
\begin{eqnarray}
\nonumber
  S=&
  \sum_{k}
  \left(\begin{array}{cc}
    \bar{\phi}_{k+} & \bar{\phi}_{k-}
  \end{array}\right)
  \left(\begin{array}{cc}
    \rmi\omega_n-\xi_{\vec{k}} & \Delta_{\rm ext}^{\phantom *} (\vec{k}) \\
    \Delta_{\rm ext}^*(\vec{k}) & \rmi\omega_n+\xi_{\vec{k}}
  \end{array}\right)
  \left(\begin{array}{c}
    \phi_{k+} \\
    \phi_{k-}
  \end{array}\right)
  \\
  &+U_0\sum_{k_1k_2k_3}
  \bar{\phi}_{k_1 +}\bar{\phi}_{k_3-k_1-k_2,-}\phi_{k_3+}\phi_{-k_2-}.
  \label{eq:hubbard_action}
\end{eqnarray}
\subsection{Propagator and effective interaction}
\begin{figure}[htbp]
  \centering
  \includegraphics{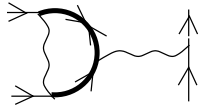}
  \caption{Diagram in $\psi$-field 
    language generating an effective interaction with three incoming 
    arrows in the symmetry-broken phase. Bold lines with one (two)
    arrow(s) represent full normal (anomalous) propagators, 
    wiggly lines represent bare interactions.}
  \label{fig:v3+1}
\end{figure}

In \eqref{eq:hubbard_action}, the external field appears
on the off-diagonal of a matrix specifying the action's 
quadratic part.
According to the Dyson equation,
inverting this matrix after subtracting the self-energy
leads to the propagator or 
one-particle Green's function
\begin{eqnarray}
  \label{eq:proint}
  \mat{G}(k)&=
  \frac{1
       }
       {\omega_n^2+E(k)^2}
  \left(\begin{array}{cc}
    -\rmi\omega_n-\xi_{\vk}-\Sigma(k) & \Delta(k) \\
    \Delta^*(k)               & -\rmi\omega_n+\xi_{\vk}+\Sigma(k)
  \end{array}\right)
  \\
  \label{eq:proabb}
  &=:
  \left(\begin{array}{cc}
    G(k) & F(k) \\
    F^*(k) & -G(-k)
  \end{array}\right),
\end{eqnarray}
where $E(k)=\sqrt{(\xi_\vk-\Sigma(k))^2+|\Delta|^2}$ and
$\Sigma(k)$ is the normal self-energy.
$\Delta$ denotes the superconducting order parameter
and contains the anomalous self-energy as well as the external field.
To simplify the diagrammatics, we antisymmetrize the interaction part 
\begin{equation}
  \label{eq:anti_hubbard_ia}
  \fl
  U_0\sum_{1\dots 4}
  \delta_{k_1+k_2,k_3+k_4}
%  \underset{(++)(--)-(+-)(-+)-(-+)(+-)+(--)(++)}{
%    \underbrace{
      \left(
        \delta_{N_1+}\delta_{N_2-}-\delta_{N_2+}\delta_{N_1-}
      \right)
      \left(
        \delta_{N_3+}\delta_{N_4-}-\delta_{N_4+}\delta_{N_3-}
      \right)
%    }
%  }
  \bar{\phi}_1
  \bar{\phi}_2
  {\phi}_3
  {\phi}_4,
\end{equation}
where $i$, if used as an index, is short for $k_i N_i$ and
$2$ and $4$ are exchanged in comparison with \eqref{eq:hubbard_action}.
%and the notation $(\pm\dots\pm):=\delta_{N_1\pm}\dots\delta_{N_m \pm}$
%has been used.
If $(N_1N_3)$ labels the rows and $(N_2N_4)$ the columns of a matrix
to the Nambu base $\{(--),(-+),(+-),(++)\}$,
the antisymmetrized bare interaction reads
\begin{equation}
  \label{eq:hubbard_ia_matrix}
  \mat{U}_0={U_0}
  \left(
    \begin{array}{cccc}
      0 & 0 & 0 & 1 \\
      0 & 0 & -1 & 0 \\
      0 & -1 & 0 & 0 \\
      1 & 0 & 0 & 0
    \end{array}
  \right)
  \delta_{k_1+k_2,k_3+k_4}.
\end{equation}
% where we have adopted the following convention
% \begin{equation}
%   \label{eq:hubbard_ia_matrix}
%   \left(
%     \begin{array}{cccc}
%       (----) & (---+) & (-+--) & (-+-+) \\
%       =(--)(--) & =(--)(-+) & =(--)(+-) & =(--)(++) \\[.15cm]
%       (--+-) & (--++) & (-++-) & (-+++) \\
%       =(-+)(--) & =(-+)(-+) & =(-+)(+-) & =(-+)(++) \\[.15cm]
%       (+---) & (+--+) & (++--) & (++-+) \\
%       =(+-)(--) & =(+-)(-+) & =(+-)(+-) & =(+-)(++) \\[.15cm]
%       (+-+-) & (+-++) & (+++-) & (++++) \\
%       =(++)(--) & =(++)(-+) & =(++)(+-) & =(++)(++)
%     \end{array}
%   \right)
% \end{equation}
By employing the Nambu field substitution 
\eqref{eq:actsub}, a formalism containing four potentially different normal 
(particle-number conserving) and twelve potentially different anomalous 
(particle-number non-conserving) effective interactions arises.
Here, as we concentrate on conventional $s$-wave pairing, we  
require time reversal invariance to hold. 
This implies the following
behavior of an effective interaction component
under flipping of all its Nambu indices:
if an odd number of the Nambu indices are $(+)$,
the sign of the effective interaction's real part is flipped;
otherwise, the sign of the imaginary part is flipped.
Exploiting this symmetry,
we assign symbols to the various effective interactions:
\begin{equation}
  \label{eq:Hubbard_U_simp_matrix}
  {\mathbb U}=
  \left(
  \begin{array}{cccc}
    X^* & \Omega^{\phantom\star}_3 & \Omega_2^{\phantom\star} & U \\
    -\Omega_4^* & W^* & V & \Omega_1 \\
    -\Omega_1^* & V^* & W & \Omega_4 \\
    U^* & -\Omega_2^* & -\Omega_3^* & X
  \end{array}
  \right).
\end{equation}
The $\Omega_i$ are collectively referred to as $3+1$ effective
interactions as they have three incoming and one outgoing leg or vice versa
if drawn according to the $\psi$-field representation of \eqref{eq:actionpsi}.
They are generated in the symmetry-broken phase by diagrams 
of the type in figure~\ref{fig:v3+1}.
$W$ is a $4+0$ anomalous effective interaction arising
only in the symmetry-broken phase.
$V$, $X$, and $U$ are normal $2+2$ effective interactions.
Only $V$ and $U$ have non-zero bare values.

\section{Mean-field and generalized RPA}
\label{sec:mf_resum}
Here we employ a mean-field approximation 
of the self-energy and a generalized random-phase approximation (RPA)
of the effective interactions. 
This represents a useful first step toward the fRG treatment, 
as it allows us to assess the 
basic behavior and the frequency- and momentum-dependence of normal 
and anomalous vertices. In particular, we find that anomalous 3+1 
vertices have a sizable feedback on 
{ certain effective interactions.}
%the interaction vertices associated 
%with the Goldstone boson. 
%Hence they should not be neglected.  

\subsection{Gap equation}

\begin{figure}[htbp]
  \centering
  \includegraphics{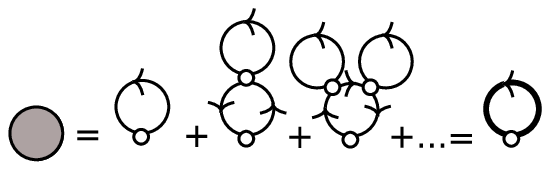}
  \caption{Perturbation expansion for the self-energy. 
    Small circles represent bare interactions, 
    lines with arrows represent bare propagators,
    bold lines with arrows represent full propagators,
    and filled circles represent the self-energy.}
  \label{fig:selper}
\end{figure}

The system's self-energy can be calculated approximatively
by resumming a subset of perturbation theory diagrams as
in figure~\ref{fig:selper}.
This subset contains all diagrams corresponding to 
bubbles, sunrise diagrams, and trees of either or both.
It is equivalent to self-consistent Hartree-Fock
theory.
If we were to restrict the Hubbard Hamiltonian to only include
the Cooper channel and forward scattering processes with zero
momentum transfer, the resulting
diagrammatic self-consistency equation would
be exact. Analogous statements can be made about the one-loop fRG treatment for such a restricted Hamiltonian\cite{2005PThPh.113.1145H}. 
The self-energy equation can be expanded into two distinct parts,
\begin{eqnarray*}
\fl
%  \label{eq:selequ}
  \Sigma&=-U_0\sum_k
  \frac{e^{\rmi\omega_n0^+}(-\rmi\omega_n-\xi_{\vec{k}}-\Sigma(\vk))+
        e^{\rmi\omega_n0^-}(\rmi\omega_n-\xi_{\vec{k}}-\Sigma(\vk))}
       {\omega_n^2+E(\vk)^2}
  \\ \fl
  \Delta-\Delta_{\rm ext}&=
  U_0  \sum_k 
  \frac{\Delta(\vk)}
       {\omega_n^2+E(\vk)^2}.
%  \label{eq:scgequ}
\end{eqnarray*}
The Matsubara
sums are performed analytically.
The result is
\begin{eqnarray}
  \label{eq:selres}
  \Sigma(\vp)&=-U_0\sum_\vk 
  \left[
    1-\frac{\xi_{\vk}+\Sigma(\vk)}
           {E_\vk}\tanh(\beta E_\vk/2)
  \right]
\\
  \label{eq:scgres}
  \Delta(\vp)&=\Delta_{\rm ext}(\vp)+
  U_0  \sum_{\vec{k}}\frac{\Delta(\vk)\tanh(\beta E_\vk/2)}
  {2E_\vk}.
\end{eqnarray}
In this approximation, both $\Sigma$ and $\Delta$ are constant in momentum 
space if the external field is.
The Hartree contribution $\Sigma$ can be absorbed into the chemical
potential and is subsequently ignored.

\subsection{Bethe-Salpeter equation}

\begin{figure}[htbp]
  \centering
  \includegraphics{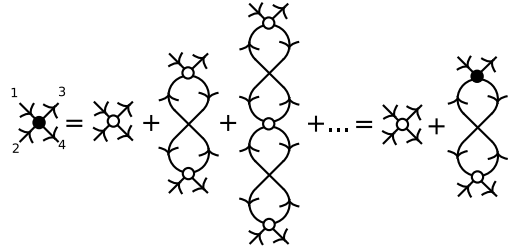}
  \caption{Bethe-Salpeter equation for the four-point function. 
    Filled circles represent effective four-point functions, 
    empty circles represent bare four-point functions, 
    lines which have both ends terminating in a vertex 
    represent full propagators. Note that the first two
    indices refer to incoming lines, while the
    second pair refers to outgoing lines.}
  \label{fig:betdia}
\end{figure}

A subset of all diagrams of the perturbation expansion for the
effective interaction can be resummed into a Bethe-Salpeter
equation as shown in figure \ref{fig:betdia}.
This corresponds to a generalization of the random-phase approximation
in the sense that particle-particle
ladders in the $\psi$-representation are included
in addition to the usual particle-hole bubbles.
Symbolically, this Bethe-Salpeter equation reads
\begin{eqnarray}\fl \nonumber
  {\mathbb U}(1234)
  =
  {\mathbb U}_0(1234)
  -
  \underset{M_1\dots M_4}{\sum_k}&
  {\mathbb U}_{N_1M_2N_3M_1}(p_1, k, p_3)
  {\mathbb U}_{0\,M_4N_2M_3N_4}(k+p_4-p_2, p_2, k)\\ \fl
  &\times{\mathbb G}_{M_1 M_4}(k+p_4-p_2)
  {\mathbb G}_{M_3 M_2}(k),
  \label{eq:hubbard_bs_symb}
\end{eqnarray}
where, as a parameter, $i=1\dots 4$ is short for $p_iN_i$.
By the introduction of the loop matrix
\begin{equation}
  \label{eq:loopmatrix}
  {\mathbb L}_{M_2 M_1, M_4 M_3}(k,p)
  :=
  {\mathbb G}_{M_1 M_4}(k+p)
  {\mathbb G}_{M_3 M_2}(k)
\end{equation}
and by exploiting momentum and energy conservation
$p_1+p_2=p_3+p_4$ while setting $p=(\vp,{\rm i}\nu_m):=p_1-p_3$, 
\eqref{eq:hubbard_bs_symb} can be rewritten as
\begin{equation}
  \label{eq:hubbard_bs_matrix}
  {\mathbb U}(p_1\dots p_4) = {\mathbb U}_0-\sum_k 
  {\mathbb U}(p_1,k,p_3,k+p)
  {\mathbb L}(k,p){\mathbb U}_0,
\end{equation}
where the Nambu indices of $\mat{U}$ and $\mat{U}_0$
are organized into pairs as in \eqref{eq:hubbard_ia_matrix}.
${\mathbb U}(p_1\dots p_4)$ only depends on $p$
in this approximation, as can be seen by
solving for $\mat{U}$,
which yields
\begin{equation}
  \label{eq:hubbard_bs_solved}
  {\mathbb U}(p)=
  \left(
    {\mathbb U_0^{-1}}
    +\sum_k {\mathbb L}(k,p)
  \right)^{-1}.
\end{equation}
% Using the symbols from \eqref{eq:proint}, 
% \begin{equation}
%   \label{eq:hubbard_loopmatrix}
%   {\mathbb L}
%   =
%   \left(
%   \begin{array}{cccc}
%     HH & HF^* & FH & FF^* \\
%     F^* H & F^* F^* & GH & GF^* \\
%     HF & HG & FF & FG \\
%     F^* F & F^* G & GF & GG 
%   \end{array}
%   \right),
% \end{equation}
% where for each matrix entry it is understood that
% $k+p$ is plugged into the first and $k$ into the
% second factor, implying that the order of the two
% propagators is fixed in this case. 
% Furthermore, a star in \eqref{eq:hubbard_loopmatrix}
% implies that the complex conjugate be taken, but only of the gap amplitude
% in the numerator of the corresponding propagator.

\subsection{Numerical setup}

The Matsubara sums in \eqref{eq:hubbard_bs_solved} can be performed
analytically. 
The integration
over the Brillouin zone and the matrix inversion on the right-hand side of
\eqref{eq:hubbard_bs_solved} is accomplished numerically.
In the following, a system with
$t^\prime = t/6$, $\mu=-1.5t$, $U_0=2t$, 
and $T=0.035t$ is assumed.
This choice of parameters corresponds to an approximately quarter-filled
system far below its mean-field critical temperature. 
$\Delta$ is chosen to be real.
We mostly employ an external field $\Delta_{\rm ext}=10^{-4}t$. 
This keeps the numerics well-behaved while having no appreciable influence 
on $\mat{U}(\vp)$ except near $\vp=0$.

\subsection{$3+1$ effective interactions}
\begin{figure}[htbp]
  \centering
  \includegraphics[scale=.38]{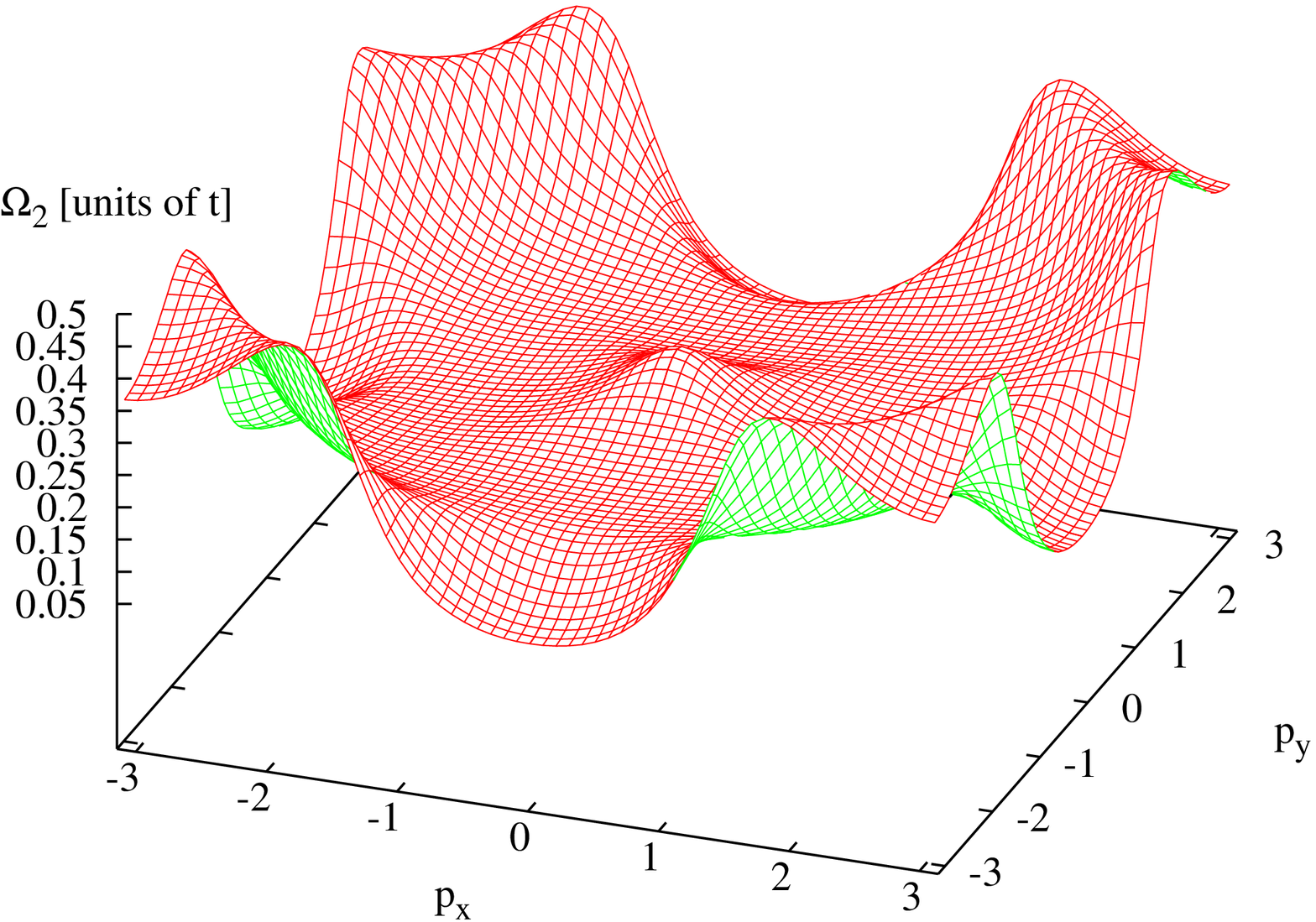}
  \includegraphics[scale=.38]{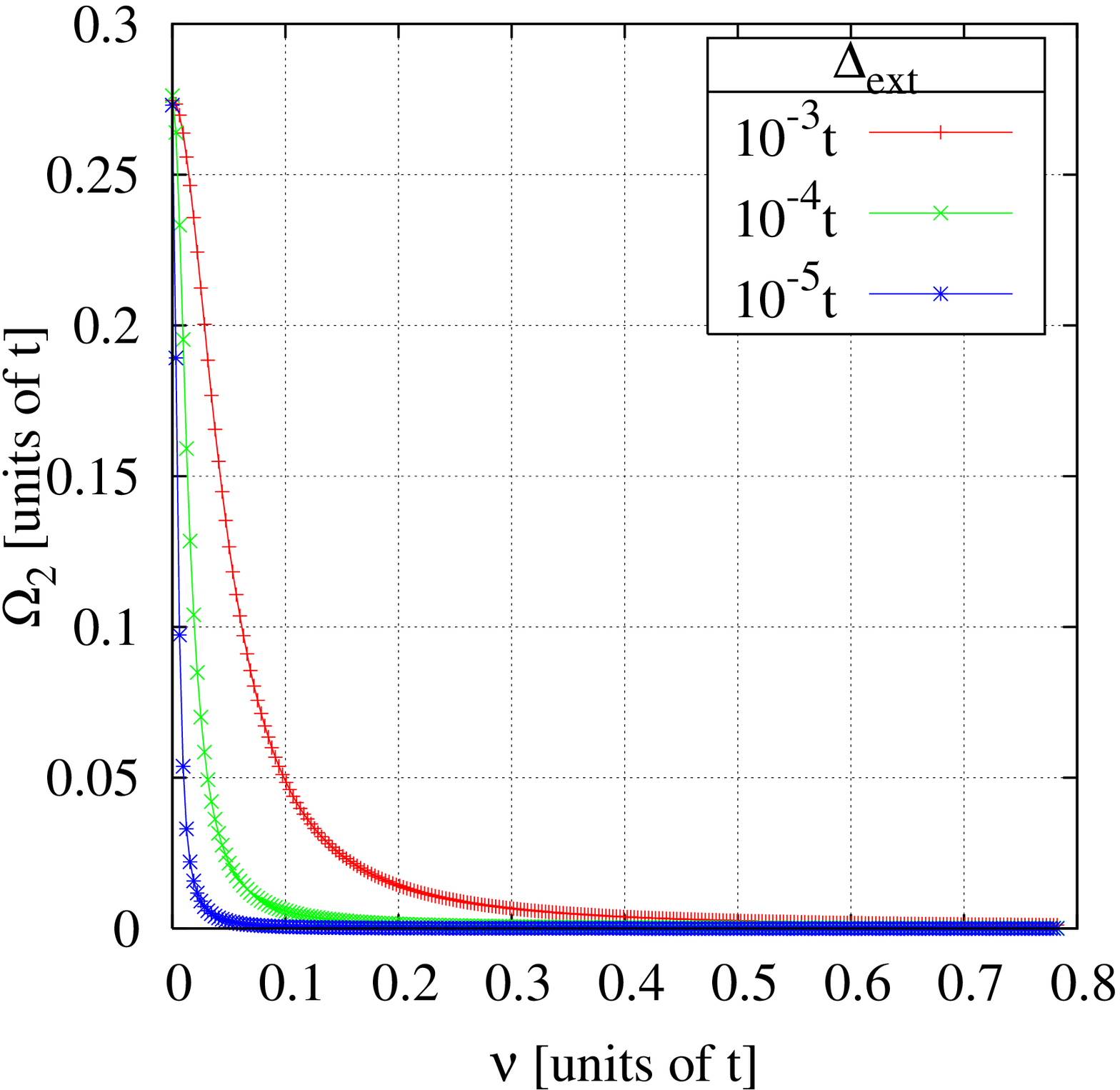}
  \caption{$\Omega_2$ for zero energy transfer 
    (left, $\Delta_{\rm ext}=10^{-4}t$) and
    zero momentum transfer (right, various $\Delta_{\rm ext}$), approximately
    quarter filling, $U_0=2t$, $\mu=-1.5t$, $t^\prime=t/6$, 
    $T<<T_{\rm c}$. $\Omega_{1,3,4}$ are similar.}
  \label{fig:Omega2}
\end{figure}

The $\Omega_i$ are generated by diagrams as in figure~\ref{fig:v3+1}. 
In the reduced mean-field models studied previously 
these vertices remain zero due to the restricted momentum-structure 
of the bare interaction where only particle pairs with 
total momentum equal to zero interact. 
However, in the general case with scattering of particle 
pairs with arbitrary total momentum they have to be considered.   
Due to the appearance of an anomalous propagator  figure~\ref{fig:v3+1},
these diagrams and hence the $\Omega_i$ are non-zero only 
in the symmetry-broken phase.
To obtain an overview of the momentum and frequency 
dependence of the $3+1$ effective
interactions, we study the graphs in figure~\ref{fig:Omega2}. 
The right-hand graphs show that the 
$3+1$ effective interactions exhibit no divergence in the 
{$\vp_1-\vp_3=0$ channel.}
The $\Omega_i$ at larger frequencies are suppressed
by decreasing the external field.
From the figure's left-hand diagram, we learn that the $\Omega_i$'s 
moduli are large around $\vp=0$ and around $\vp=(\pi,\pi)$
for $\nu=0$.

\subsection{Large interaction vertices and Goldstone boson}

The instability toward superconductivity gives rise to a growth of certain 
combinations of the interaction vertices. 
From previous fRG studies on reduced mean-field models
\cite{2004PThPh.112..943S,2005EPJB...48..349G} and from the solution 
of the Bethe-Salpeter equation, the mean-field picture of the leading 
components for zero total incoming momentum and frequency 
has been obtained as follows.
{There is}
one strongly growing combination of effective interactions that occurs 
in the flow equation for the anomalous self-energy. 
It typically 
reaches its maximum at the critical scale or temperature $T_{\rm c}$ 
and does not diverge below. 
In the BCS mean-field model, this combination 
is given by $V+W$ for real $\Delta$ \cite{2004PThPh.112..943S}. 
Another combination of the effective
interactions, $V-W$, diverges also below  $T_{\rm c}$.
This divergence is associated with the massless phase mode dubbed
Goldstone boson present if a continuous symmetry, in our case a 
global $U(1)$-symmetry, is broken.
For cases where no continuous symmetry is broken as for the half-filled 
charge-density-wave mean-field model with a two-fold degenerate ground 
state \cite{2005EPJB...48..349G}, the cutoff-dependence of the effective 
interaction is comparable to the cutoff dependence of the combination 
$V+W$ of normal and anomalous effective interactions
of a BCS mean-field model.

For the Hubbard interaction studied in this paper, 
the decay of these large vertices away from zero-total-momentum or 
-frequency becomes important, as other channels not present in the 
mean-field models might pick up large contributions from the 
symmetry-breaking channel.  
In this section, we investigate $V+W$ and $V-W$ with regard
to their momentum and energy dependence as well as to the
impact of the $3+1$ effective interactions.
 
\subsubsection{Momentum dependence}
\label{sec:momdep}
\begin{figure}[htbp]
  \centering
  \includegraphics[scale=.37]{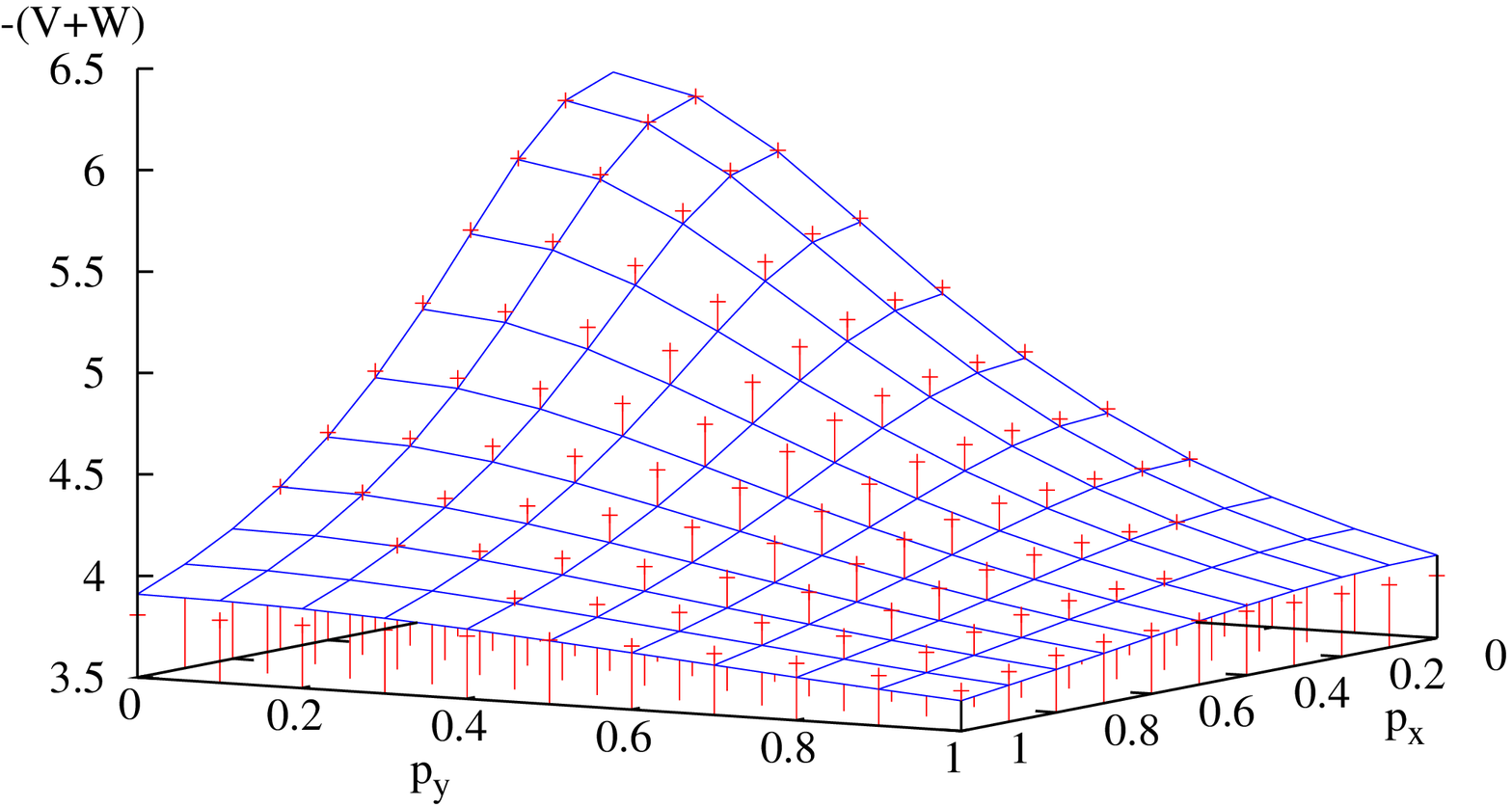}
  \includegraphics[scale=.37]{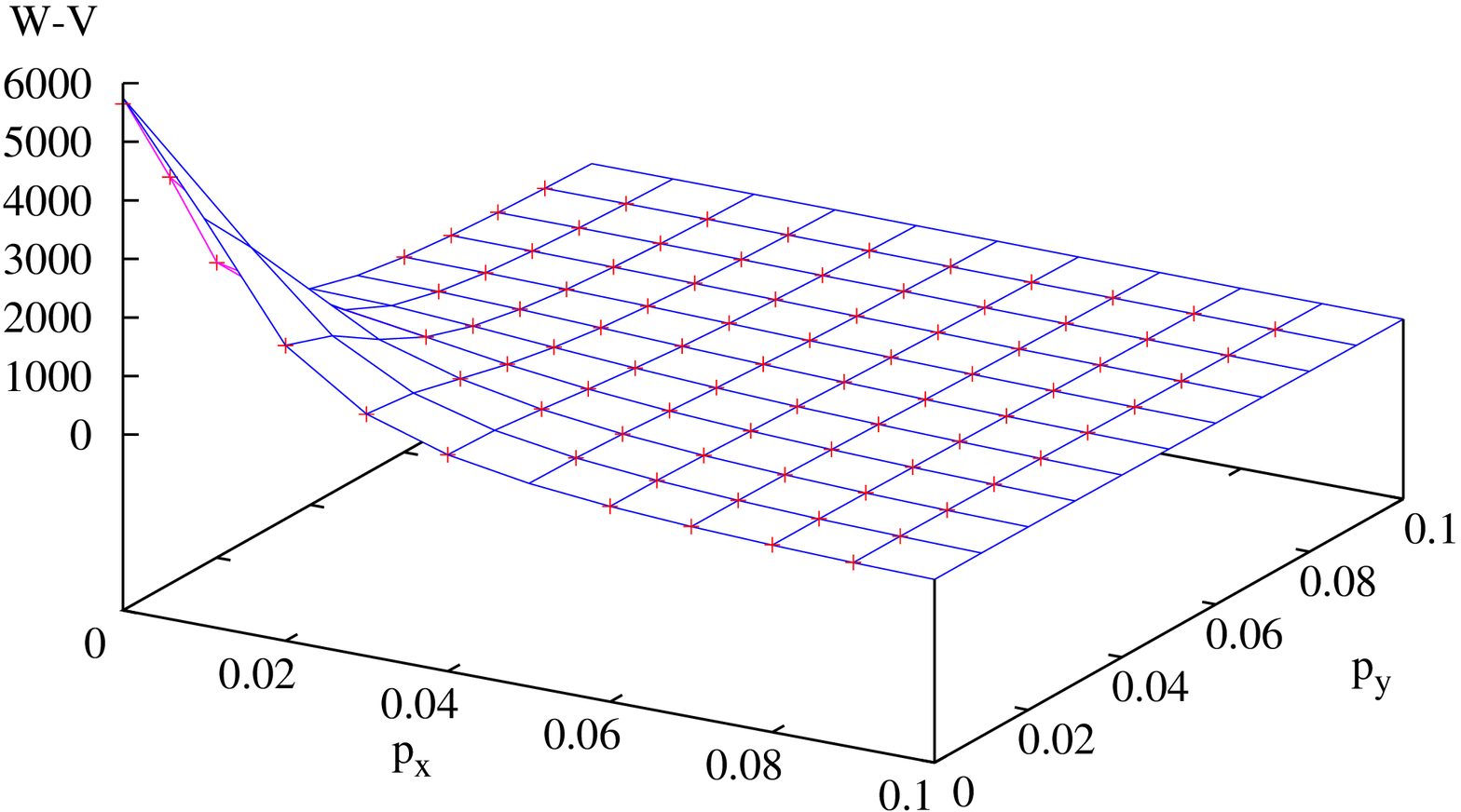}
  \caption{Left: Blue lines represent \eqref{eq:vertex_comb_approx} 
    fitted to $-(V+W)$ obtained from numerics (red crosses). Both plotted
     in units of $t$, 
     $A\approx 0.37t^{-1}$, 
% %    $\alpha=-0.369405t^2$, 
     $\beta\approx 1.32t^{-1}$, 
     $\gamma\approx 3.32t$. 
      Right: Blue lines represent 
      \eqref{eq:vertex_comb_approx} fitted to $W-V$ obtained
      from numerics (red crosses). Both plotted 
      in units of $t$,
% % %    $\alpha=-0.5743$, 
      $A\approx 1.8t^{-2}\Delta_{\rm ext}$,
% % %    $\alpha\approx-0.57t^2$, 
      $\beta\approx 0.81t^{-1}$, $\gamma\approx 3.33t$.
    Both diagrams: $U_0=2t$, $\mu=-1.5t$, $t^\prime=t/6$, 
      $T<<T_{\rm c}$, $\Delta_{\rm ext}=10^{-4}t$.}
  \label{fig:pV+Wsquared}
\end{figure}

We analyze the momentum dependence of $V-W$ and $V+W$
at $\nu_m=0$.
We fit 
\begin{equation}
  \label{eq:vertex_comb_approx}
  \frac{1}{\beta |\vec{p}|^2 + A}+\gamma,
\end{equation}
to data obtained from numerical calculations 
for small up to intermediate $\vp$,
see figure~\ref{fig:pV+Wsquared}.
For the non-divergent part $V+W$,
the fit reproduces the numerically-calculated behavior.
For small momenta, the agreement is better,
while for larger momenta, deviations become larger,
but vanish along the axes.
This is due to an anisotropy in the numerical data which
is not captured by the approximation \eqref{eq:vertex_comb_approx}.

For the divergent part
$V-W$, we note that $A$ as obtained from the fitting procedure
is proportional to $\Delta_{\rm ext}$.
This is confirmed by calculations for $\Delta_{\rm ext}=10^{-3}t$
and $\Delta_{\rm ext}=10^{-5}t$.
% , where $A\approx 1.7t^{-2}\Delta_{\rm ext}$ 
% and $A=??$, respectively. 
For intermediate momenta along the diagonal (not shown), 
the fit is not optimal as deviations of up to $25\%$ occur. 
The situation is better for
large momenta, where deviations of about $10\%$ are common.
The peak at $\vp=0$ is well-captured by the approximation
\eqref{eq:vertex_comb_approx}, which permits us to 
confirm the dispersion relation for the low-lying 
Goldstone mode in the next section.

\begin{figure}[htbp]
  \centering
  \includegraphics[scale=.37]{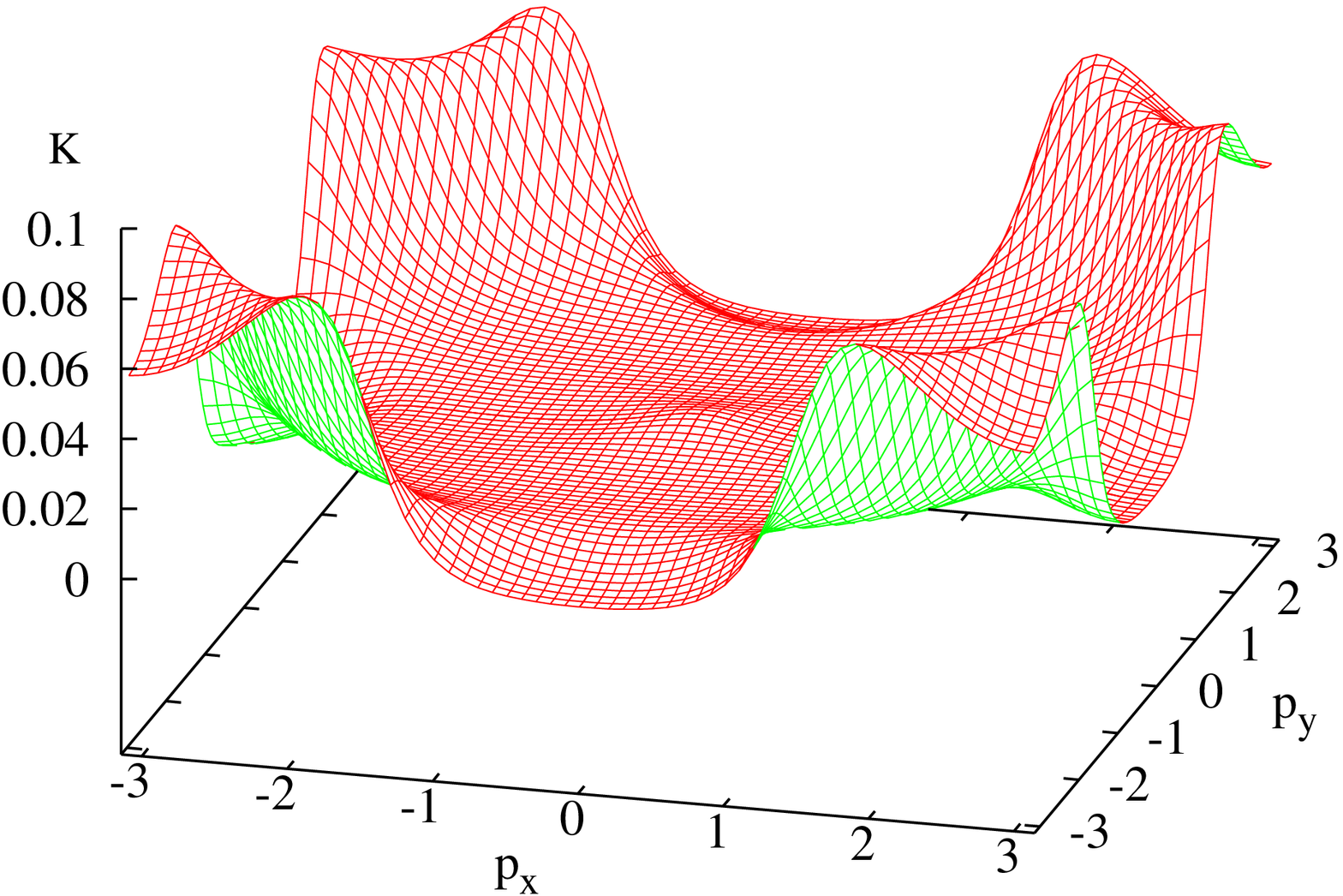}
  \includegraphics[scale=.37]{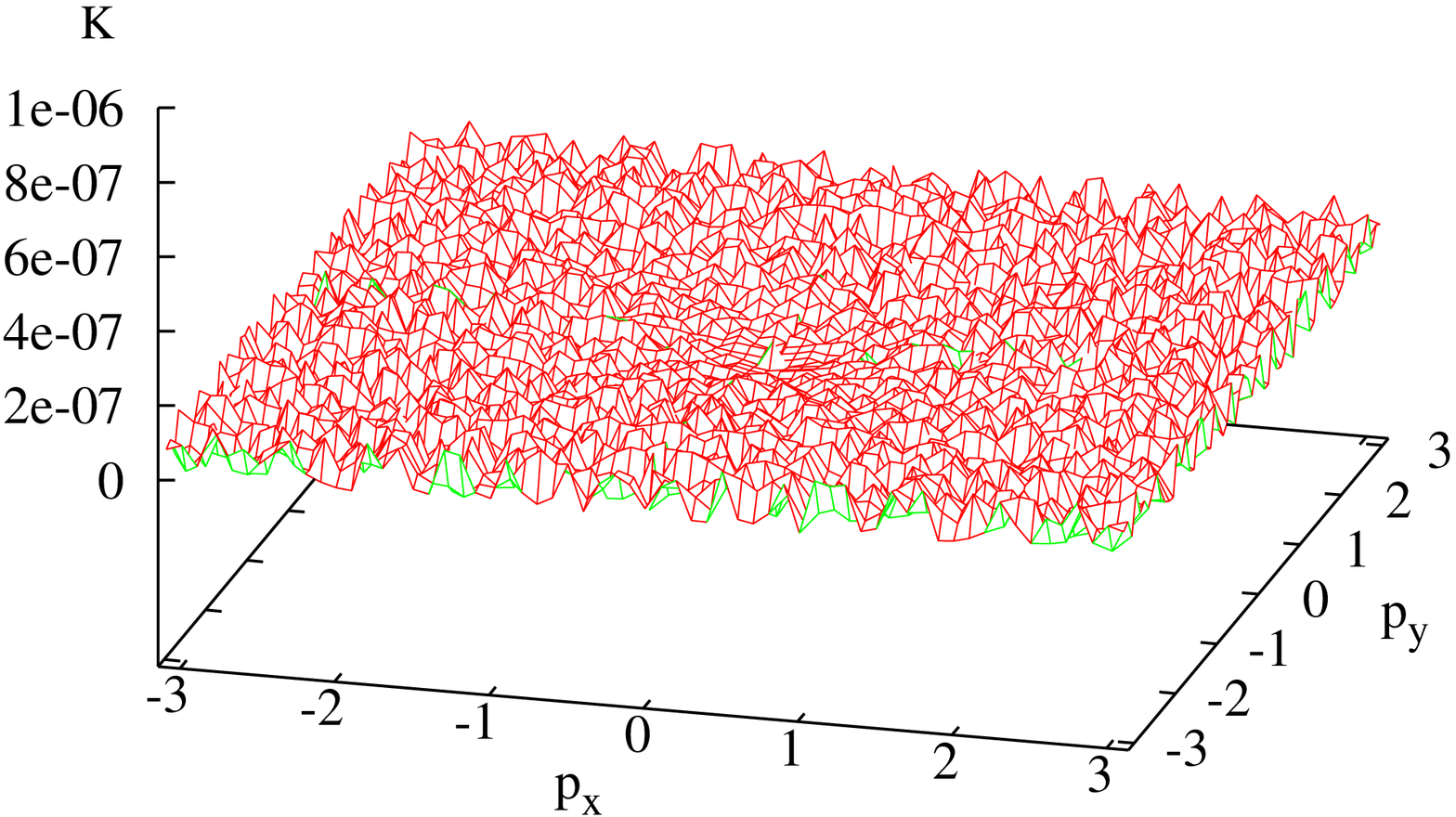}
  \caption{Modulus of the relative change of $V+W$ (left)
  and $V-W$ (right) upon considering $3+1$ effective interactions.
  Both diagrams: $U_0=2t$, $\mu=-1.5t$, $t^\prime=t/6$, 
  $T<<T_{\rm c}$, $\Delta_{\rm ext}=10^{-4}t$}
  \label{fig:modes_change_p}
\end{figure}

We consider the impact of the $3+1$ effective interactions.
By setting loops combining an anomalous and a normal propagator
to zero in \eqref{eq:hubbard_bs_symb}, the $3+1$ effective
interactions are set to zero.
The modulus 
\begin{equation}
  K:=\left|\frac{\Gamma_{\rm w/o\,\,3+1}-\Gamma_{\rm with\,\,3+1}}
       {\Gamma_{\rm w/o\,\,3+1}}\right|
     \label{eq:Gammachange}
\end{equation}
of the relative change of a given effective interaction
combination $\Gamma$ upon switching on the $3+1$ effective interactions
is plotted for $\Gamma=V+W$ and $\Gamma=V-W$
in figure~\ref{fig:modes_change_p}.
We note that the magnitude of this change for the combination $V+W$ 
largely follows the magnitude of the $3+1$ effective interactions
in the Brillouin zone.
It varies between $0$ and $10\%$.
The combination $V-W$ corresponding to the system's phase mode
does not change to numerical accuracy (note the scale on the $z$-axis).

\subsubsection{Energy dependence}

\begin{figure}[htbp]
  \centering
  \includegraphics[scale=.3]{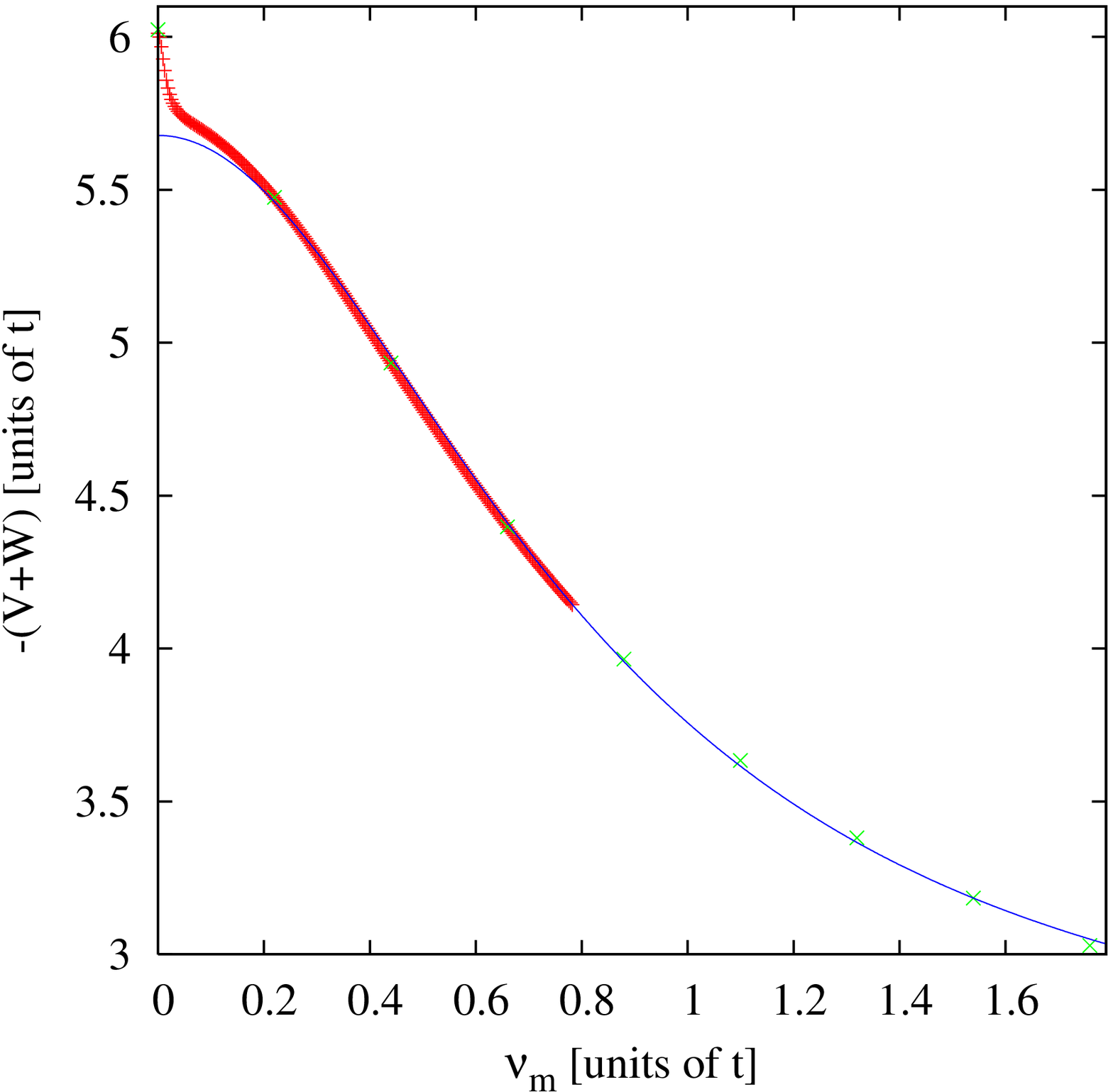}
  \includegraphics[scale=.3]{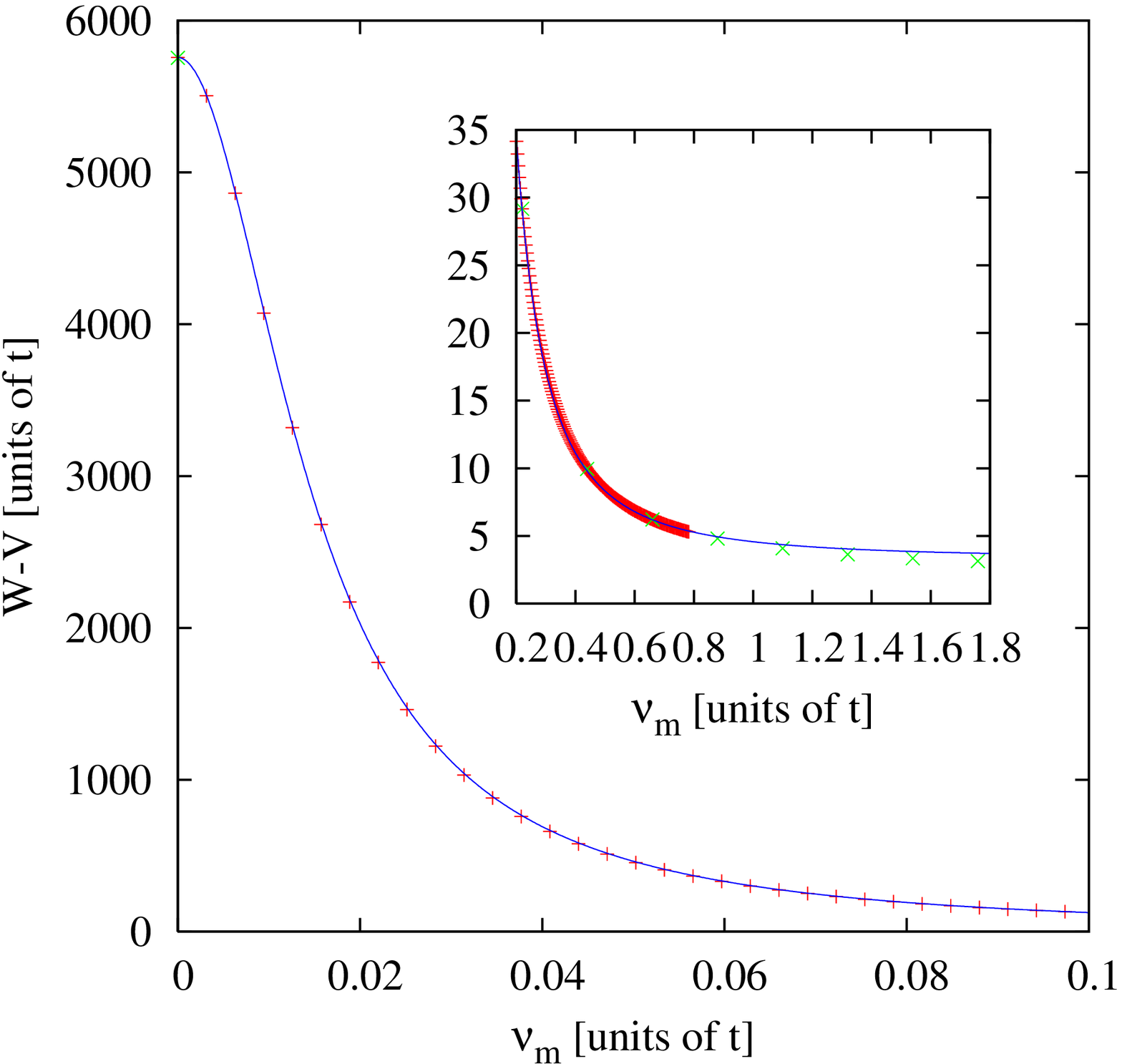}
  \caption{Matsubara-frequency dependence of $-(V+W)$ (left)
    and $W-V$ (right) calculated 
    far below $T_{\rm c}$ (green crosses: low resolution, 
    red crosses: high resolution). The blue lines are least-squares fits to
    \eqref{eq:vertex_comb_approx_V+W_en}.
%     \eqref{eq:vertex_comb_approx_V+W_en_alt},
%     \eqref{eq:vertex_comb_approx_V+W_en_alt2}.
  The parameters used in the fits are, from left to right, 
  $\{\beta_\nu\approx 0.47t^{-3}, \alpha\approx 0.31t^{-1}, 
  \gamma\approx 2.49t\}$,
  $\{\alpha\approx-0.58t^{-1}$,
    $\beta_\nu\approx0.46t^{-3}$, $\gamma\approx-3.33t\}$.
    Both diagrams: $U_0=2t$, $\mu=-1.5t$, $t^\prime=t/6$, 
    $T<<T_{\rm c}$, $\Delta_{\rm ext}=10^{-4}t$}

%   $\{\beta_\nu=-0.40t^{-1}, \alpha=-0.29t, \gamma=-2.31t, 
%   \alpha_{\exp}=0.24t, \beta_{\exp}=3030t^{-2}\}$,
%   $\{\beta_\nu=-0.37t^{-1}, \alpha=-0.28t, \gamma=-2.23t, r=1.72 \} $}
  \label{fig:nu_V+W_squared}
\end{figure}

The energy dependence of $V+W$ and $V-W$ is to be included
in the approximate formula \eqref{eq:vertex_comb_approx}
by adding $\beta_\nu \nu_m^2$ to the denominator, i.e.
by considering the approximation function
\begin{equation}
  \label{eq:vertex_comb_approx_V+W_en}
  \frac{1}{\beta_p |\vec{p}|^2 + \beta_\nu \nu_m^2 + \alpha}+\gamma.
\end{equation}
We restrict ourselves to the real part of the effective interactions
for brevity.
Contemplating the left-hand graph of
figure \ref{fig:nu_V+W_squared},
we become aware of a shoulder close to zero frequency,
which is not captured by
\eqref{eq:vertex_comb_approx_V+W_en}.
This implies a small non-linearity of the
dispersion relation of the amplitude mode.
% We employ the alternative approximation function
% \begin{equation}
%   \label{eq:vertex_comb_approx_V+W_en_alt}
%   \frac{t^2}{\beta_p |\vec{p}|^2 + \beta_\nu \nu_m^{1.8} + \alpha}+\gamma
%   -\alpha_{\rm exp}e^{-\beta_{\rm exp}\nu_m^2}
% \end{equation}
% As seen from
% the middle graph of figure \ref{fig:nu_V+W_squared}, the numerical
% data is reproduced including the shoulder.
% In order to reduce the number of 
% free parameters compared to \eqref{eq:vertex_comb_approx_V+W_en_alt}
% but improve the fit compared to \eqref{eq:vertex_comb_approx_V+W_en},
% we study the third alternative
% \begin{equation}
%   \label{eq:vertex_comb_approx_V+W_en_alt2}
%   \frac{t^2}{\beta_p |\vec{p}|^2 + \beta_\nu \nu_m^r + \alpha}+\gamma,
% \end{equation}
% which is expanded in comparison to \eqref{eq:vertex_approx_en}
% and \eqref{eq:vertex_comb_approx_V+W_en}
% in that the exponent of $\nu_m$ in the denominator is not fixed.
% This yields the right-most graph. Again, the important value
% at $\nu_m=0$ is underestimated, albeit less so than in the left-hand
% plot. 
% We note that the best fits
% \eqref{eq:vertex_comb_approx_V+W_en_alt} and 
% \eqref{eq:vertex_comb_approx_V+W_en_alt2}
% are obtained for an exponent of $\nu_m$ different
% from $2$ in the denominator. As the exponent of
% $|\vp|$ was $2$, this implies a small change
% in the linearity of the dispersion of the amplitude mode.

The ansatz \eqref{eq:vertex_comb_approx_V+W_en} 
reproduces the numerically-obtained results for $V-W$
upon properly adjusting the parameters,
as the right-hand part of figure~\ref{fig:nu_V+W_squared} illustrates. 
The agreement is good for small Matsubara frequencies. 
In the inset, we see that the large-frequency behavior is not captured
optimally, which is due to our choice of $\gamma$, having
taken it from the fit in figure~\ref{fig:pV+Wsquared}. 
However, the exponent $2$ for $\nu_m$ is confirmed 
by the fit.
This together with the momentum-dependence fit in
section \ref{sec:momdep} confirms the linearity of the dispersion relation of
the Goldstone mode connected to $V-W$.

\begin{figure}[htbp]
  \centering
  \includegraphics[scale=.3]{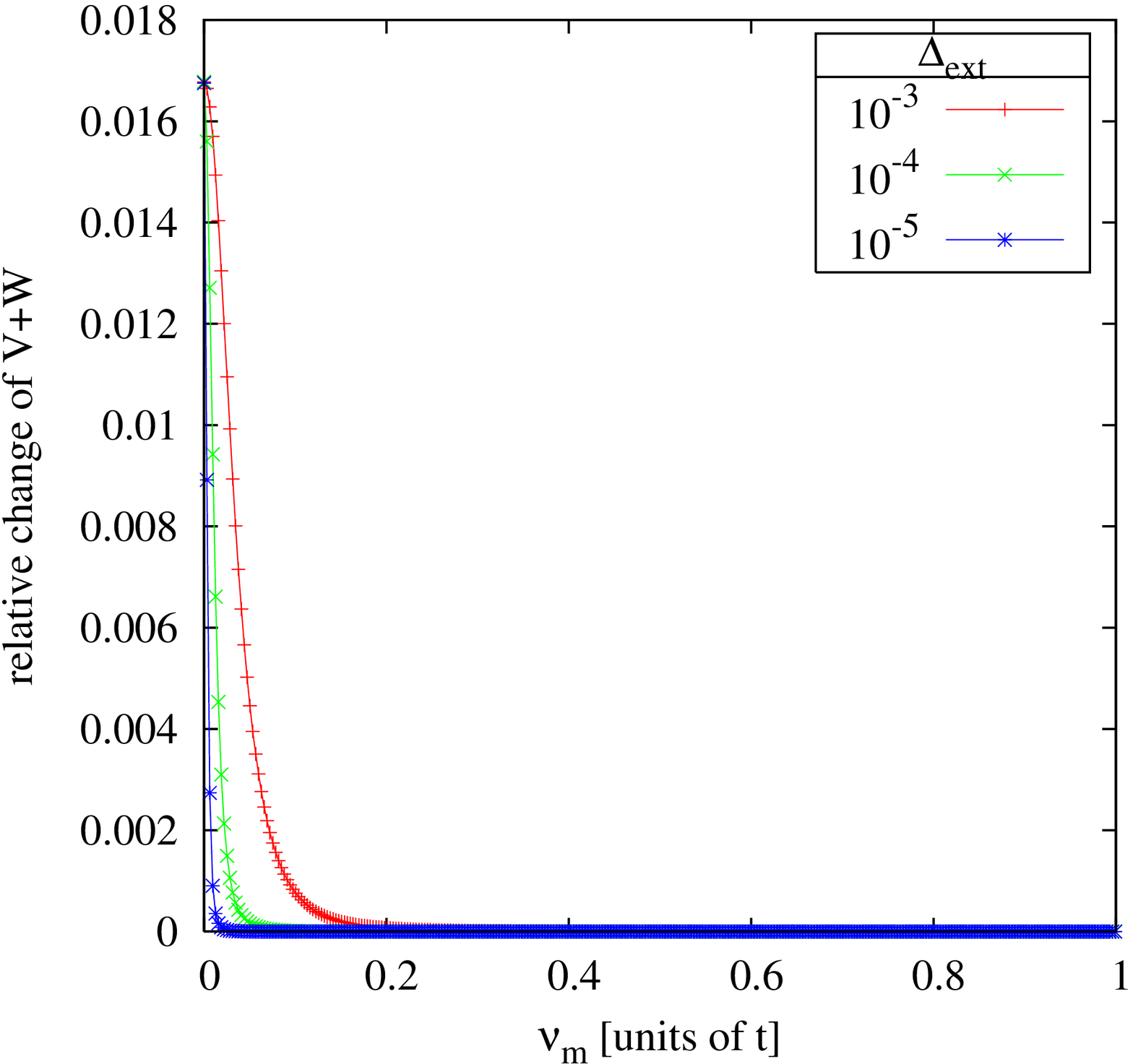}
  \includegraphics[scale=.3]{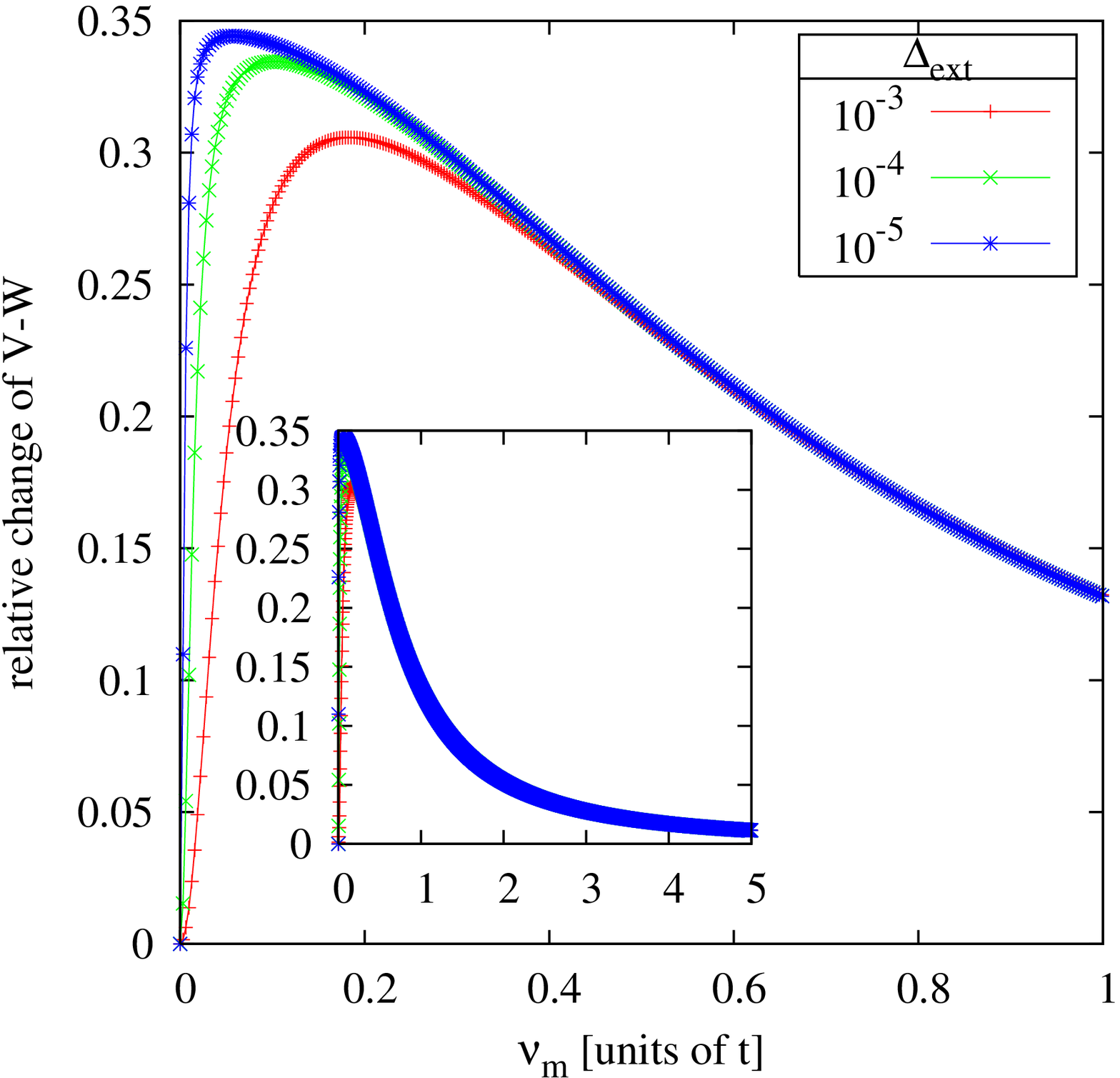}
  \caption{Relative change of the energy dependence of 
    $V+W$ (left) and $V-W$ (right) upon considering the $\Omega_i$.
    Both diagrams: $U_0=2t$, $\mu=-1.5t$, $t^\prime=t/6$, 
    $T<<T_{\rm c}$, $\Delta_{\rm ext}=10^{-4}t$}
  \label{fig:nu_comb_change}
\end{figure}

To understand the influence of the $\Omega_i$ ($i=1\dots 4$) 
on the energy dependence of $V+W$ and $V-W$,
we plot the relative change at $\vp=0$ as in \eqref{eq:Gammachange}, 
see figure~\ref{fig:nu_comb_change}.
We glean that for $V+W$, the only appreciable change
occurs at $\nu_m=0$ and is quite small in the limit $\Delta_{\rm ext}\to 0$.
For $V-W$, the converse is true: Only at $\nu_m=0$ is
there no change, while even for small $\nu_m>0$, a sizable change
is observed which is only suppressed at large $\nu_m$.
This implies that it would not be prudent to neglect
the $\Omega_i$, while the smallness of their influence
on the $\nu_m=\vp=0$ part of $V+W$ suggests  that their
quantitative impact on the superconducting gap obtained by an
fRG calculation is likely small.
The latter supposition is motivated by the observation that
$V+W$ drives the flow of the gap, as already shown for the BCS mean-field model
in \cite{2004PThPh.112..943S}.

\section{Functional renormalization group}
\label{sec:frg}
In this section, 
we describe the fRG treatment of the attractive Hubbard model on 
the two-dimensional square lattice {at zero temperature. 
Our main goals are to study the flows produced 
as well as to obtain reliable quantitative results for the order parameter.}
Compared to the partial resummation 
in the last section, the fRG takes into account all one-loop diagrams. 
{The consequential
increase in numerical complexity is balanced by adopting
a cruder discretization of momentum and frequency space.
By taking into account all one-loop diagrams,
the pairing channel is renormalized by other 
processes with non-zero total momentum such as spin- and charge fluctuations. 
}
We will find that this causes a reduction of the superconducting order 
parameter with respect to the mean-field result. 
Furthermore, the fRG treatment allows for a $\vec{k}$-dependence 
of the self-energy around the Fermi surface.  
\subsection{Momentum-shell and interaction-flow procedures}
We introduce a cutoff $\Lambda$ into our action
by substituting 
\begin{equation}
  \G_0 \to \chi(\Lambda)\G_0.
  \label{eq:cutoff}
\end{equation}
% For each component of $\Q$, there is a component of $\chi$, and
% the division is to be understood component-wise.
We call
\begin{equation}
  \label{eq:momcutoff}
  \chi(\Lambda)=\Theta(|\xi_\vk| - \Lambda)
\end{equation}
{\em momentum-shell} cutoff function \cite{PhysRevB.4.3184} and
\begin{equation}
  \label{eq:iacutoff}
  \chi(\Lambda)=\sqrt{\Lambda}
\end{equation}
{\em interaction-flow} cutoff function \cite{2004PhRvB..70w5115H}.
The former name is reflected 
in the expansion of the system by an infinitesimal shell in momentum space 
if $\Lambda\rightarrow \Lambda+\rmd\Lambda$.
The latter name can be understood by scaling the fields $\psi$
in \eqref{eq:actionpsi} to $\sqrt[4]{\Lambda}\psi$, causing the
non-interacting part of the action 
to be cutoff-independent and the interacting,
quartic part to be proportional to $\Lambda$. 
For the momentum-shell flow,
setting $\Lambda=\Lambda_{\rm i}>\max_{\vk}{|\xi_\vk|}$
provides a solvable, non-interacting starting point.
For $\Lambda=\Lambda_{\rm f}=0$, the original, fully interacting
system is recovered.
For the interaction flow, $\Lambda=\Lambda_{\rm i}=0$
suppresses the interaction according to 
the rescaling of the fields outlined above 
and provides a solvable starting point.
$\Lambda=\Lambda_{\rm f}=1$ recovers the original, fully interacting
system.

Employing the momentum-shell cutoff, we introduce a small
external field $\Delta_{\rm ext}$ to select the symmetry-broken
phase as the endpoint of the fRG calculation.
This means our results in this case apply to a situation with
spontaneously-broken symmetry only in the limit $\Delta_{\rm ext}\to 0$.
For the interaction-flow cutoff, we add zero in
the form $\Delta_{\rm ext}-\Delta_{\rm ext}$ to the quadratic
part of the action, including $\Delta_{\rm ext}$ in the non-interacting
part and $-\Delta_{\rm ext}$ in the interacting part of the action.
This approach has been shown to 
yield symmetry-broken results at zero external field
\cite{2006NJPh....8..320G} for $\Lambda=\Lambda_{\rm f}$.
% Previous implementations of flows with physical cutoffs
% had the calculational advantage of yielding a relevant result
% for each $\Lambda$
% \cite{2004PhRvB..70w5115H,PhysRevB.64.184516}.

\subsection{fRG equations in the Katanin truncation}
\begin{figure}[htbp]
  \centering
  \includegraphics{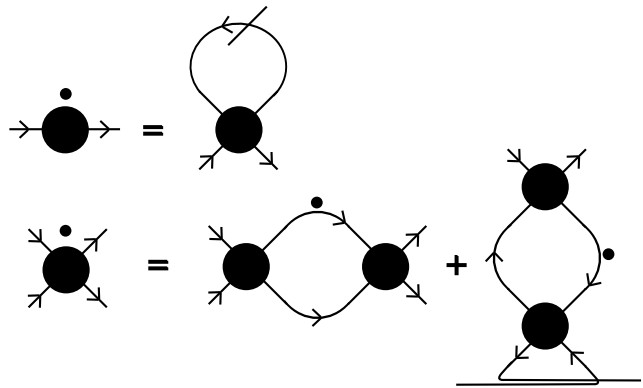}
  \caption{1PI flow equations in the Katanin truncation.
    Internal lines stand for full propagators,
    a slashed full line represents the single-scale
    propagator $\mat{S}$,
    a disc with two legs stands for the self-energy,     
    a disc with four legs represents the effective interaction,
    a dot denotes differentiation with respect to the cutoff
    $\Lambda$.
  }
  \label{fig:flow_equations}
\end{figure}

The flow equations which we employ are shown in figure~\ref{fig:flow_equations},
where the single-scale propagator 
$\mat{S}=-\G\,\partial_\Lambda\left(\G_0^{-1}\right)\G$
is used.
The name can be understood by noting that because of the differentiation
acting on $\Theta$, $\mat{S}$ lives
only on the energy scale $\Lambda$ for the momentum-shell flow.
These flow equations can be obtained by truncating according to Katanin
\cite{2004PhRvB..70k5109K} the infinite hierarchy 
of one-particle irreducible (1PI)
flow equations of \cite{2001PThPh.105....1S}.
Solving them yields the exact order parameter
and effective interaction in
the thermodynamic limit for any
model with only two-particle interactions which is mean-field exact
\cite{2005EPJB...48..349G,gersch:236,2004PThPh.112..943S}.

Symbolically, the flow equation for $\mat{U}$ reads
\begin{eqnarray}
  \label{eq:fullUflow}
\fl\nonumber
  \dot{\mat{U}}&(1234)
  =-\!\sum_{k,\vec{M}} \mat{U}_{N_1N_2M_4M_2}(p_1,p_2,k)
    \mat{U}_{M_3M_1N_3N_4}(k,-k+p,p_3)\dot{\G}_{M_3M_4}(k)\G_{M_1M_2}(-k+p)\\
\fl\nonumber
  &+\sum_{k,\vec{M}}
    \mat{U}_{M_1N_2M_4N_4}(k-q,p_2,k)
    \mat{U}_{N_1M_3N_3M_2}(p_1,k,p_3) 
    \partial_\Lambda\left[{\G}_{M_3M_4}(k)\G_{M_1M_2}(k-q)\right]
     \\ 
% \fl\nonumber & &
%     +\mat{U}_{M_3N_2M_2N_4}(k,p_2,k+q)
%     \mat{U}_{N_1M_1N_3M_4}(p_1,k+q,p_3)
%     \dot{\G}_{M_3M_4}(k)\G_{M_1M_2}(k+q)
%\\[.3cm] 
     \fl  &  -\sum_{k,\vec{M}}
    \mat{U}_{M_1N_1M_4N_4}(k-q^\prime,p_1,k)
    \mat{U}_{N_2M_3N_3M_2}(p_2,k,p_3) 
    \partial_\Lambda\left[{\G}_{M_3M_4}(k)\G_{M_1M_2}(k-q^\prime)\right],
    \label{eq:Uflow}
%     \\ 
% \fl\nonumber & &
%     +\mat{U}_{M_3N_1M_2N_4}(k,p_1,k+q^\prime)
%     \mat{U}_{N_2M_1N_3M_4}(p_2,k+q^\prime,p_3)
%     \dot{\G}_{M_3M_4}(k)\G_{M_1M_2}(k+q^\prime),
\end{eqnarray}
where $p=p_1+p_2$, $q=p_3-p_1$, and $q^\prime=p_3-p_2$.
Note that the last two terms correspond to the second diagram pictured above,
as the antisymmetrization is denoted explicitly in the symbolic equation.
Prefactors arising from Fourier transformations are absorbed
into the summation symbols.

The symbolic form of the flow equation for the self-energy reads
\begin{eqnarray}
  \dot{\Sigma}_{N_1N_2}(p)
%   &=\sum_{k,M_1M_2}
%   \mat{U}_{M_1N_2M_2N_1}(k,p,k)
%   \mat{S}_{M_1M_2}(k)\\
  =\!\sum_{k,M_1M_2}
  \mat{U}_{M_1N_2M_2N_1}(k,p,k)
  \dot{\chi}(k)\frac{\partial\G_{M_1M_2}}{\partial\chi}(k).
  \label{eq:sigmaflow}
\end{eqnarray}
Here, the single-scale propagator is rewritten 
into a form better suited for numerical evaluation
\cite{2005EPJB...48..349G}.

\subsection{Numerical results}
\begin{figure}[htbp]
  \centering
  \includegraphics[scale=.5]{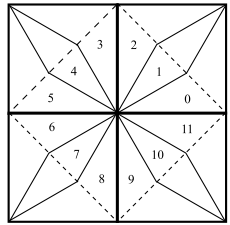}
  \caption{Patching of the Brillouin zone used with varying
    total number of patches in the numerical calculations.
    Every patch covers the same angle.}
  \label{fig:patching}
\end{figure}

In the following, we simplify \eqref{eq:Uflow} and \eqref{eq:sigmaflow}
and subsequently present results stemming from a numerical quadrature
of the simplified flow equations.
While we focus on the momentum and frequency structure
of the effective interaction in section \ref{sec:mf_resum},
in this section we are mainly interested in
determining the order parameter beyond mean-field theory, 
taking into account all one-loop diagrams in the fRG.
However, this comes at the price of an increase in 
numerical complexity, which we balance by adopting a cruder
discretization of the effective interaction.
We first drop the frequency-dependence of $\mat{U}$.
Consequently, no energy dependence of $\Sigma$ is generated in the flow
according to \eqref{eq:sigmaflow}.
We can therefore do all Matsubara sums 
analytically.
We furthermore drop the normal part of the self-energy,
thus neglecting Fermi surface shifts.
All quantities can be considered real in our approximation.
A parametrization taking into account partially the
momentum dependence of the effective interaction
is the so-called ``patching'' \cite{0295-5075-44-2-235,2000PhRvB..6113609Z}.
The Brillouin zone is split into patches as shown in 
figure~\ref{fig:patching}.
The functional value of the effective interaction
at the mid-angle of the patch and on the Fermi surface is
assumed to be also valid for all other momenta in the patch.
Thereby, only the momentum dependence along the Fermi surface is taken into
account.
This is motivated by the finding that the dependence
perpendicular to the Fermi surface is irrelevant in the
RG sense for the momentum-shell flow without symmetry-breaking
(\cite{1991PhyA..177..530S,1994RvMP...66..129S}).
We specify the momentum arguments of $\mat{U}$
as triples of patch indices, so $V(1,2,3)$ refers
to any $V(\vp_1,\vp_2,\vp_3)$ with $\vp_1$ in patch $1$, $\vp_2$ in patch $2$,
and $\vp_3$ in patch $3$.
Note that to obtain momenta in the $\psi$ representation from these
patch numbers, signs must be adjusted according to \eqref{eq:actsub}.

\subsubsection{Momentum-shell flows}
\label{sec:momentumflow}
We replace the step function from
\eqref{eq:momcutoff} by
\begin{equation}
  \label{eq:soft_cutoff}
  \chi(\Lambda)=1-\frac{1}{e^{\alpha(|\xi|-\Lambda)/\Lambda}+1}.
\end{equation}
This simplifies the numerical calculation significantly, 
although one could suspect that the sharp cutoff might
eliminate an integration. 
However, in the Katanin truncation, 
terms proportional to the derivative of the order parameter arise
in addition to the terms proportional to the derivative of the cutoff function.
Note that our results do not depend on 
$\alpha$ in a reasonable range, which adjusts the sharpness of the cutoff. 

\begin{figure}[htbp]
  \centering
  \includegraphics[scale=.3]{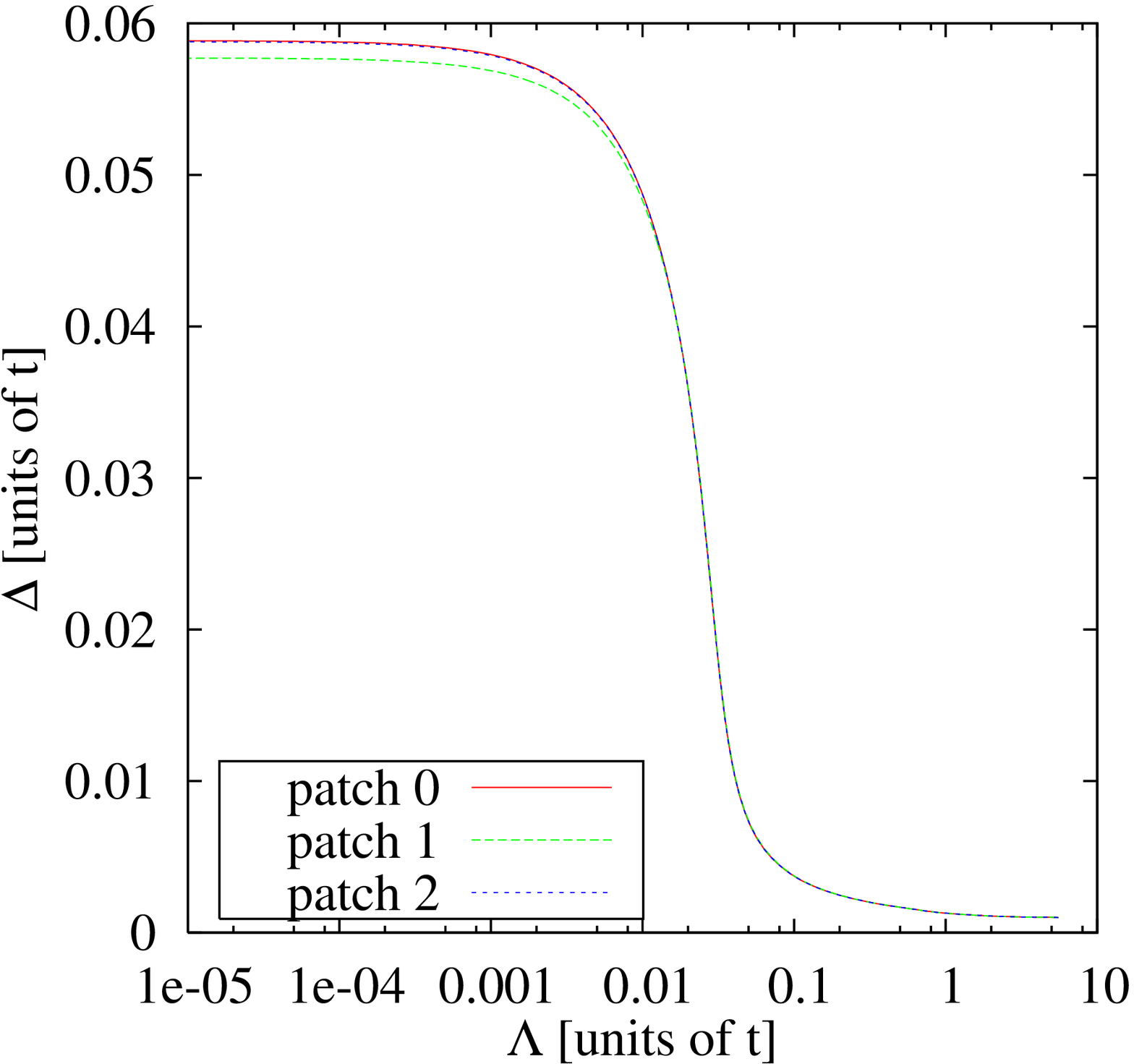}  
  \includegraphics[scale=.3]{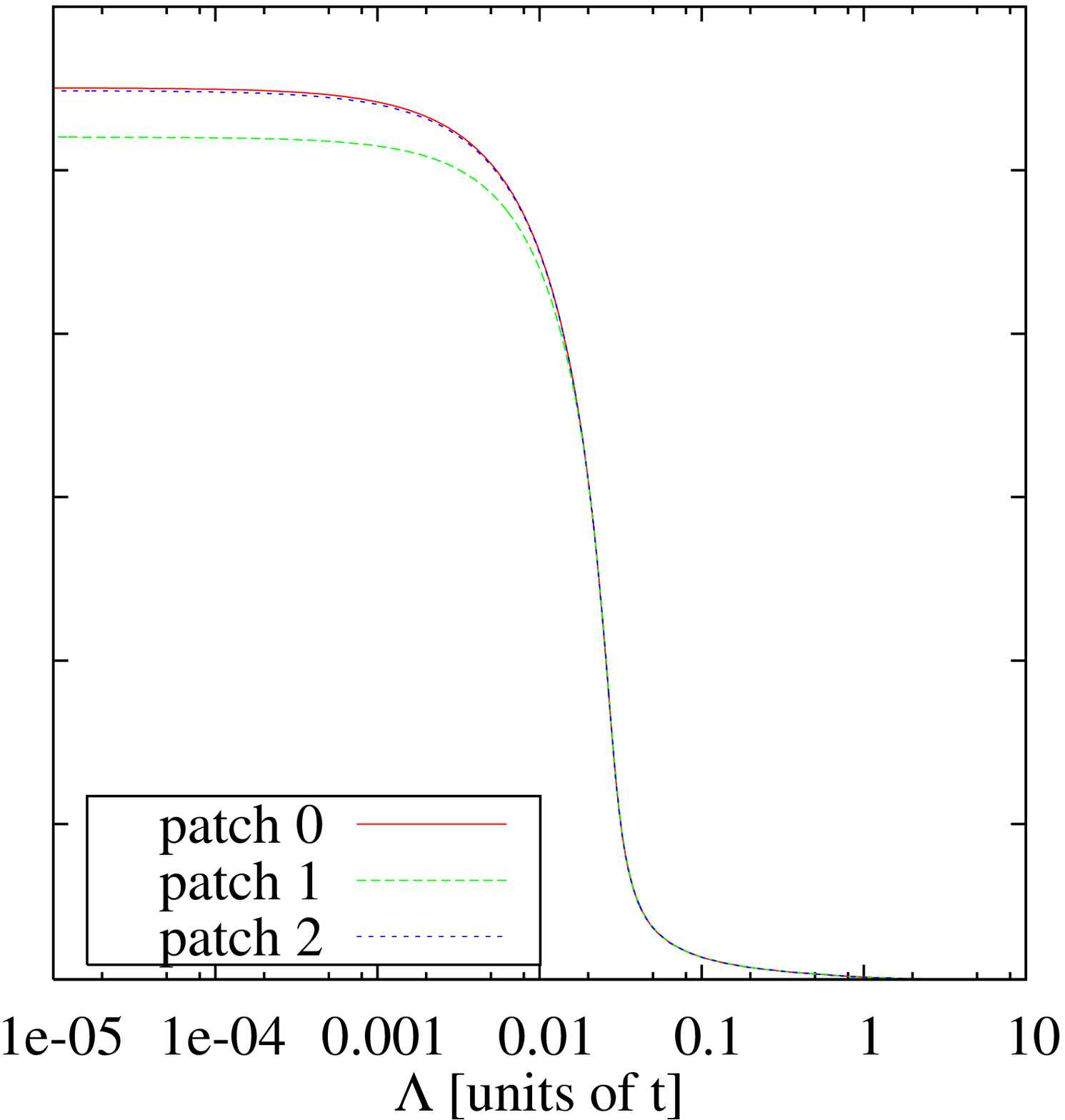}
  \caption{Momentum-shell flow of the order parameter. Twelve patches, quarter filling: $\mu=-1.41t$, $t^\prime=-0.1t$; $U_0=1.5t$, $\Delta_{\rm ext}=10^{-3}t$ (left), $\Delta_{\rm ext}=5\times10^{-4}t$ (right).}
  \label{fig:opdiverging}
\end{figure}

In figure~\ref{fig:opdiverging},
we plot flows of the order parameter in the first quadrant of
the Brillouin zone.
We note that the flow saturates for the two external
field strengths employed.
However, if we employ too small an external field, the effective
interaction flows diverge and the method breaks down.
% To illustrate this, we plot in figure~\ref{fig:divergence_flow}
% \begin{figure}[hbtp]
%   \centering
%   \includegraphics[scale=.3]{divergence_flow}
%   \caption{Magnitude of the order parameter in the flow against the 
%     Cooper-channel effective interaction in the flow, which starts at the bottom left of each graph and terminates at the top right. Twelve patches, quarter filling: $\mu=-1.41t$, $t^\prime=-0.1t$, $U_0=1.5t$, $\Delta_{\rm ext}$ varies, numbers in parentheses are patch numbers to be understood within the Nambu formalism (see text).}
%   \label{fig:divergence_flow}
% \end{figure}
% the relationship between a Cooper-channel effective coupling
% and the order parameter for various external fields.
% The momentum arguments of the 
% effective interaction plotted are specified by a combination of three patch
% numbers in parentheses.
Fortunately, the external field can be chosen at least
a hundred times smaller than the final value of the
superconducting gap.
In calculations for mean-field models \cite{2005EPJB...48..349G}, 
the modification of the order parameter 
due to the external field is less than $20\%$ in this case.

\begin{figure}[htbp]
  \centering
  \includegraphics[scale=.3]{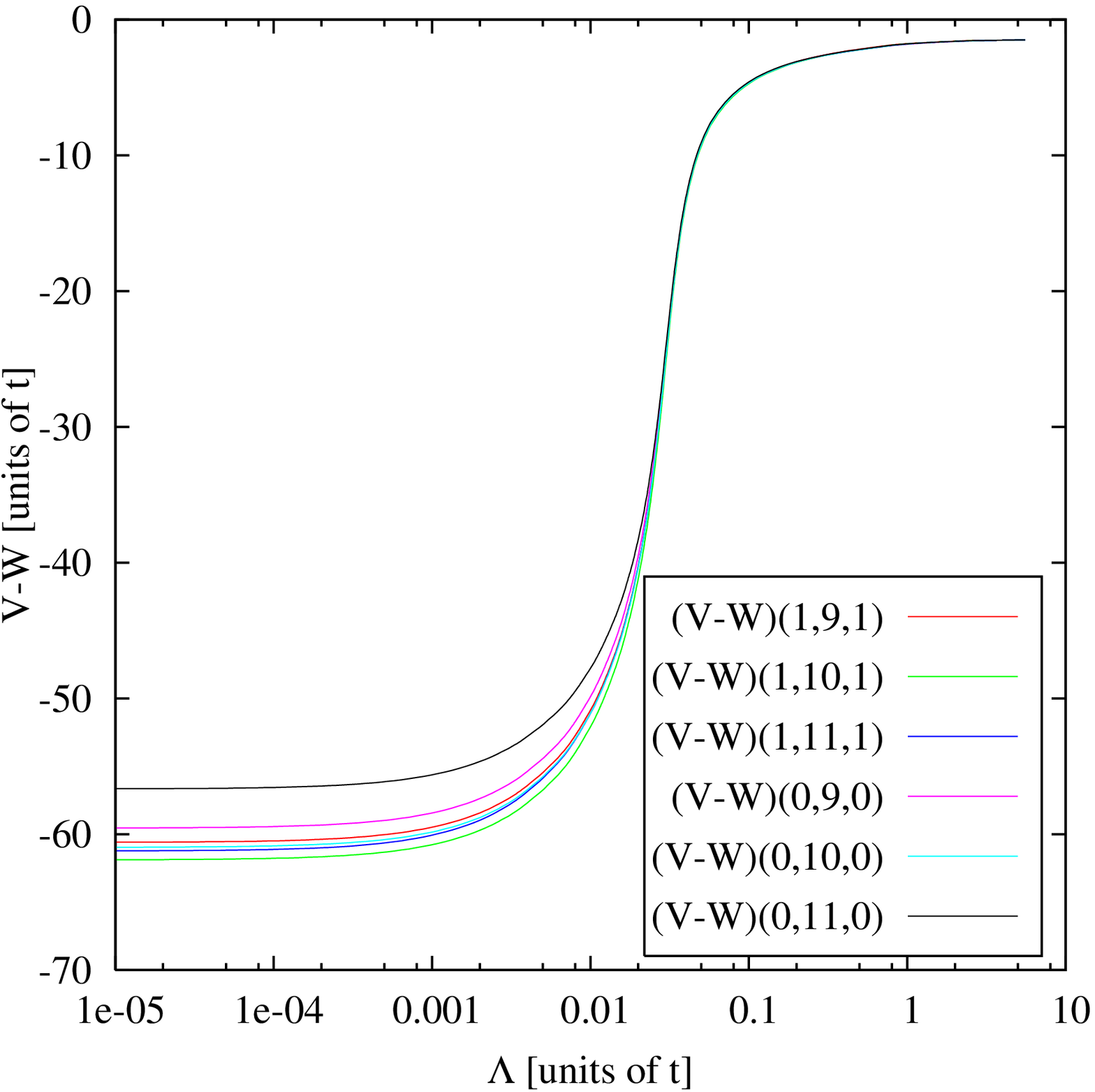}  
  \includegraphics[scale=.3]{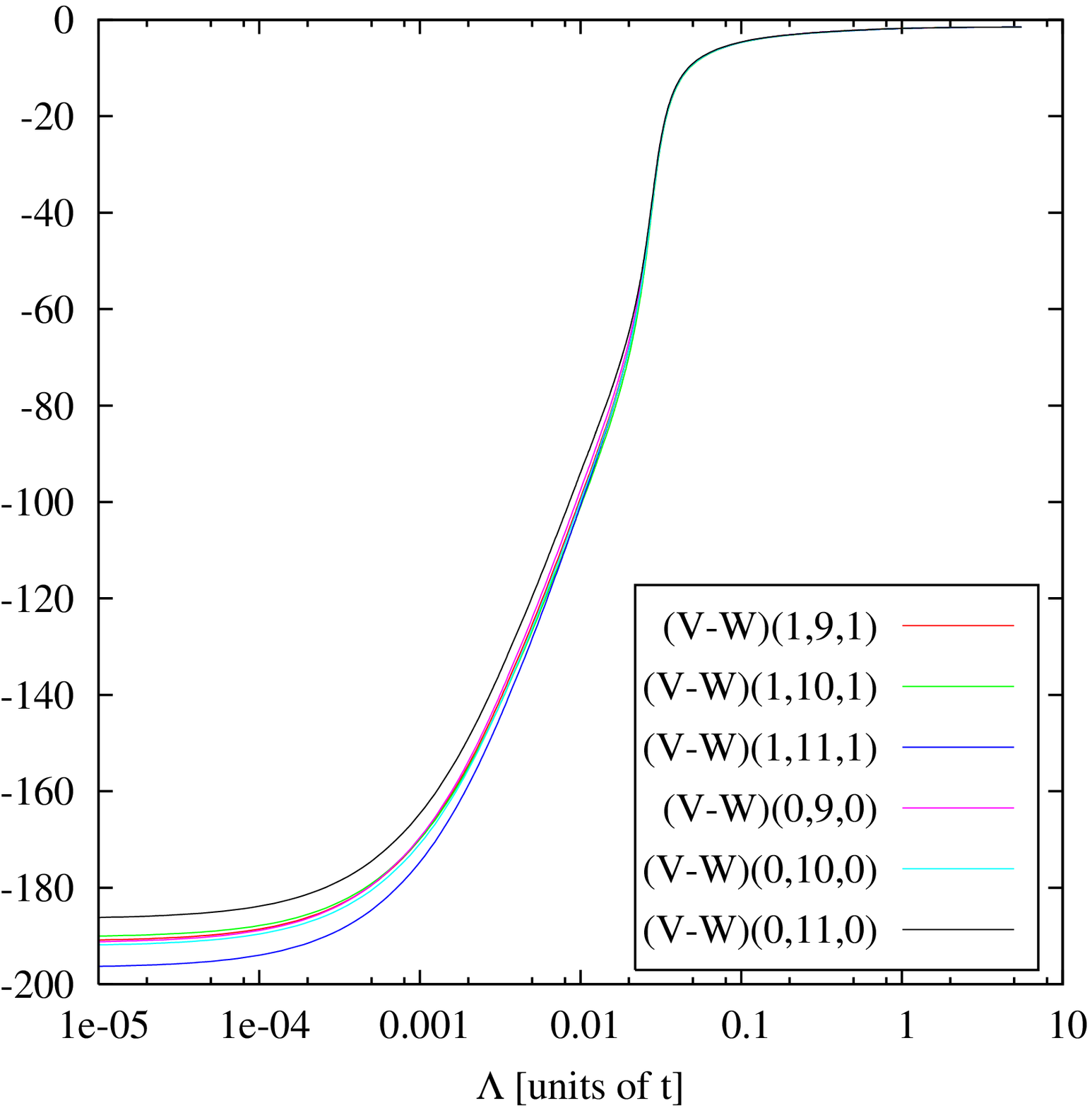}
  \caption{Momentum-shell flows of the effective interaction $V-W$ associated with the Goldstone mode. Twelve patches, quarter filling: $\mu=-1.41t$, $t^\prime=-0.1t$; $U_0=1.5t$, $\Delta_{\rm ext}=10^{-3}t$ (left), $\Delta_{\rm ext}=5\times10^{-4}t$ (right), numbers in parentheses are patch numbers to be understood within the Nambu formalism.}
  \label{fig:V-Wflow}
\end{figure}

From the resummation studies in section \ref{sec:mf_resum},
we have learned that the Goldstone phase mode $V-W$ dominates
for zero momentum transfer in the Nambu formalism, which is equivalent
to zero total momentum if we transform back to the original
$\psi$ fields according to \eqref{eq:actsub}.
In figure~\ref{fig:V-Wflow}, we plot several flows from this
channel. 
They dominate in magnitude all other effective interaction flows, including 
those not shown in the figure, and exhibit a spread of $\sim10\%$
under variation of the momentum combination.
They are quite similar to the flows found in \cite{2004PThPh.112..943S}.
Note that the maximal final value modulus increases stronger
than proportionally to $\Delta_{\rm ext}^{-1}$.
This is a precursor of the divergence to arise for even smaller
$\Delta_{\rm ext}$.

\begin{figure}[htbp]
  \centering
  \includegraphics[scale=.3]{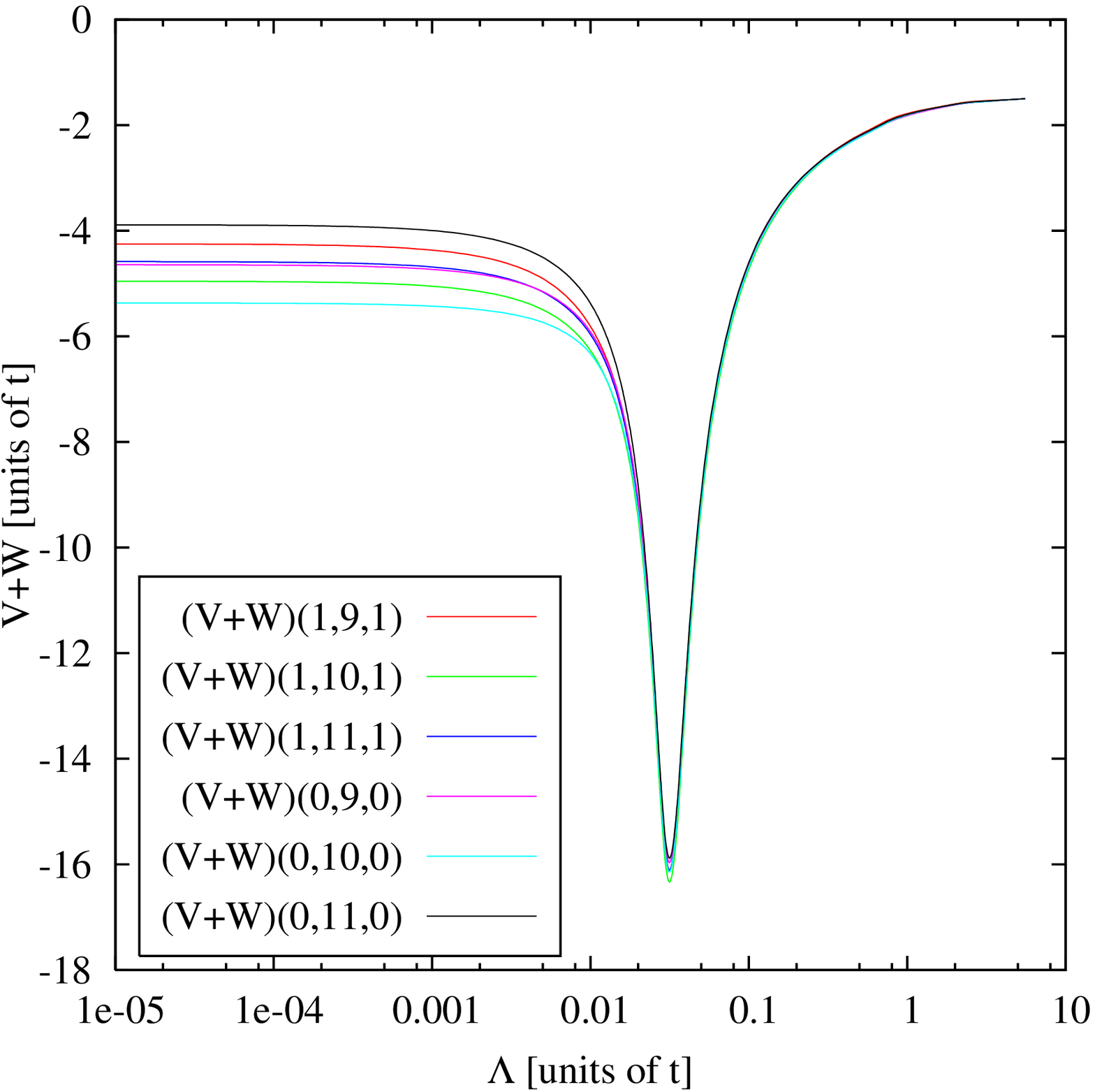}  
  \includegraphics[scale=.3]{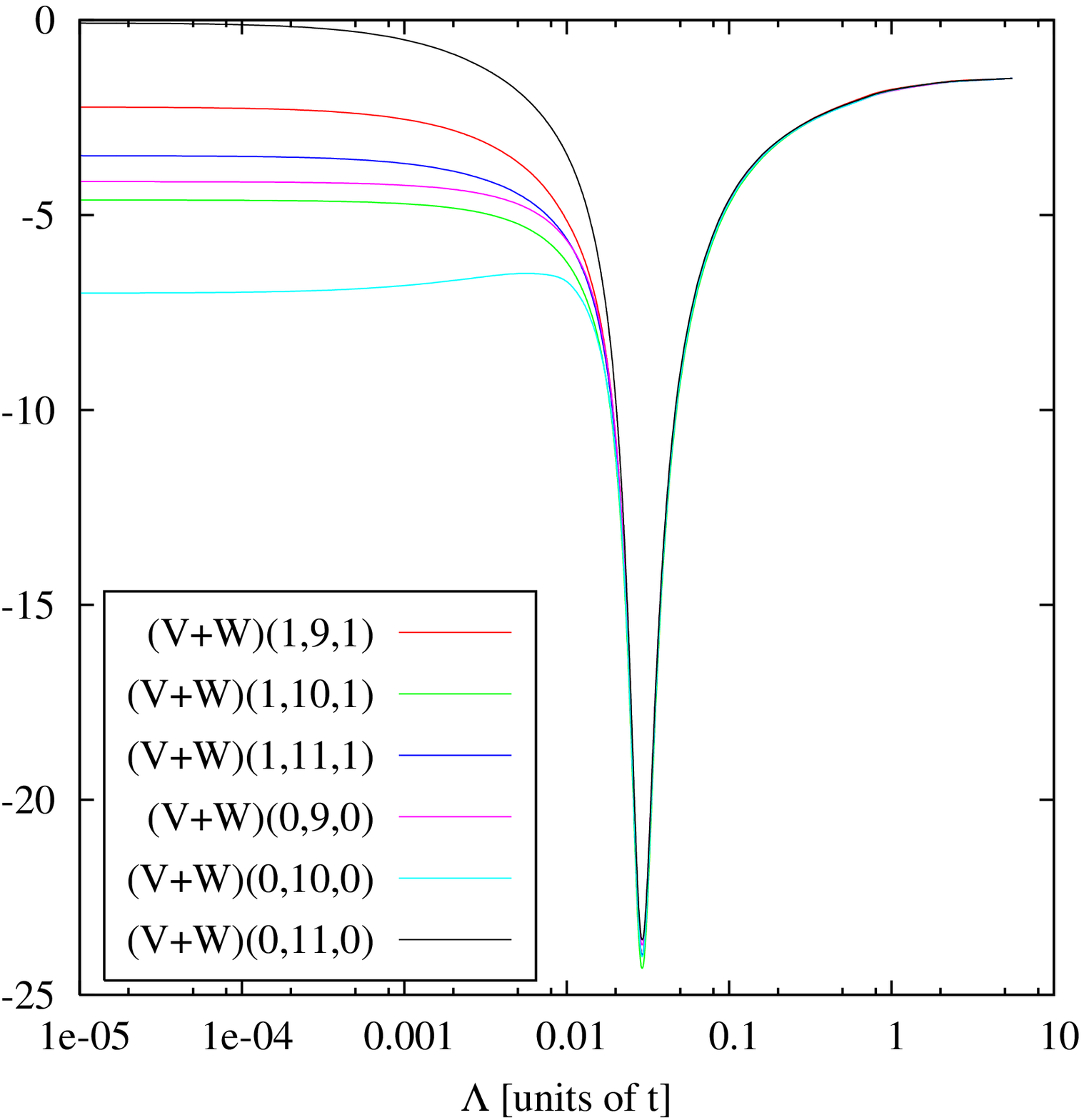}
  \caption{Momentum-shell flows of the effective interaction $V+W$ associated with the amplitude mode. Twelve patches, quarter filling: $\mu=-1.41t$, $t^\prime=-0.1t$; $U_0=1.5t$, $\Delta_{\rm ext}=10^{-3}t$ (left), $\Delta_{\rm ext}=5\times10^{-4}t$ (right), numbers in parentheses are patch numbers to be understood within the Nambu formalism.}
  \label{fig:V+Wflow}
\end{figure}

Complementing the phase mode is the effective interaction
combination $V+W$ driving the order parameter.
Flows of $V+W$ are plotted in figure~\ref{fig:V+Wflow}.
The spread of the final values increases with decreasing
$\Delta_{\rm ext}$, while the average remains approximately constant.
The graphs resemble those found when a discrete symmetry
is broken \cite{2005EPJB...48..349G}.
However, the monotonicity below the critical
scale which had been established analytically for the
mean-field model is violated by one of the flows
in the right-hand figure.
Overall, all momentum-shell flows 
exhibited resemble the flows for the mean-field
models.

\subsubsection{Interaction flows}
\begin{figure}[htbp]
  \centering
  \includegraphics[scale=.3]{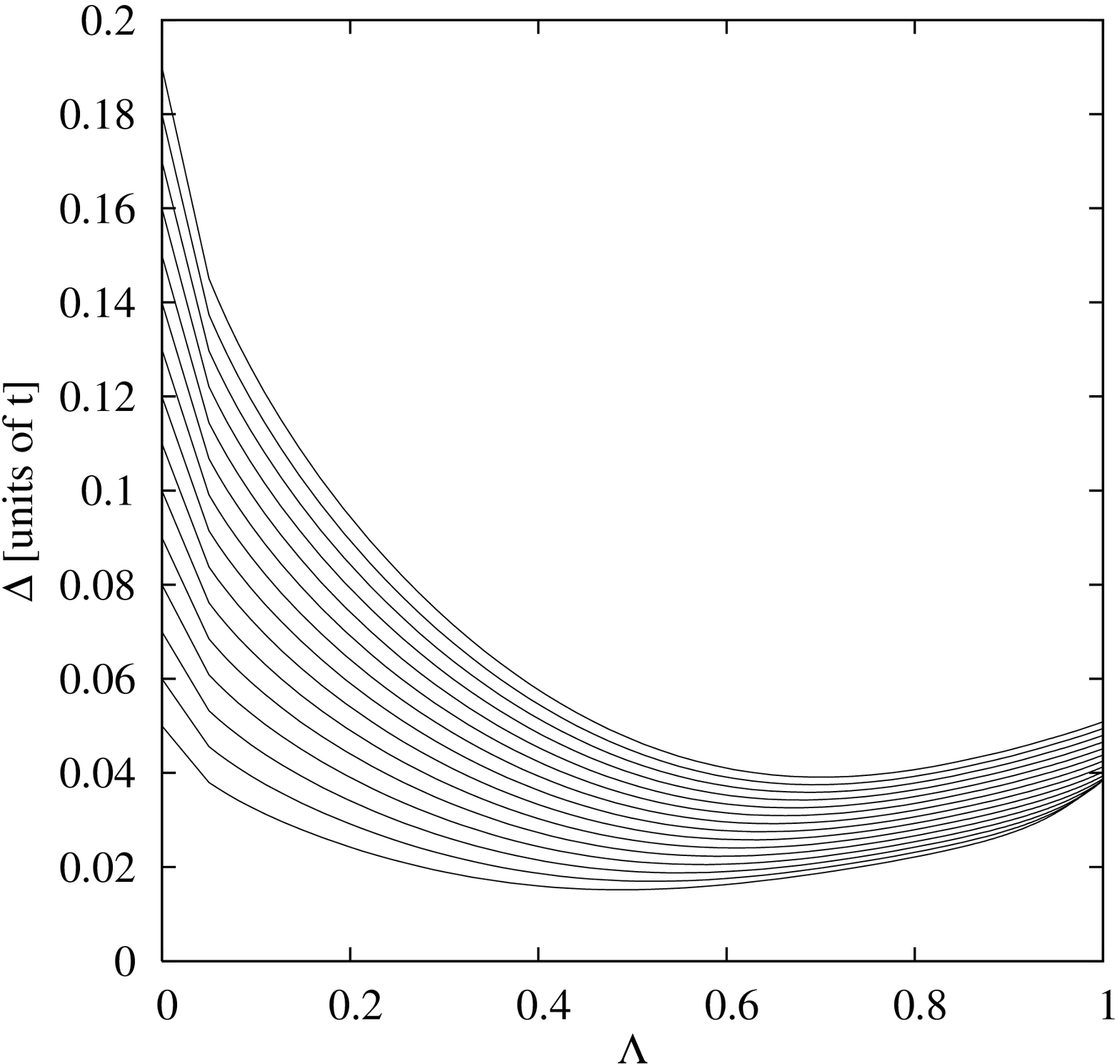}
  \includegraphics[scale=.3]{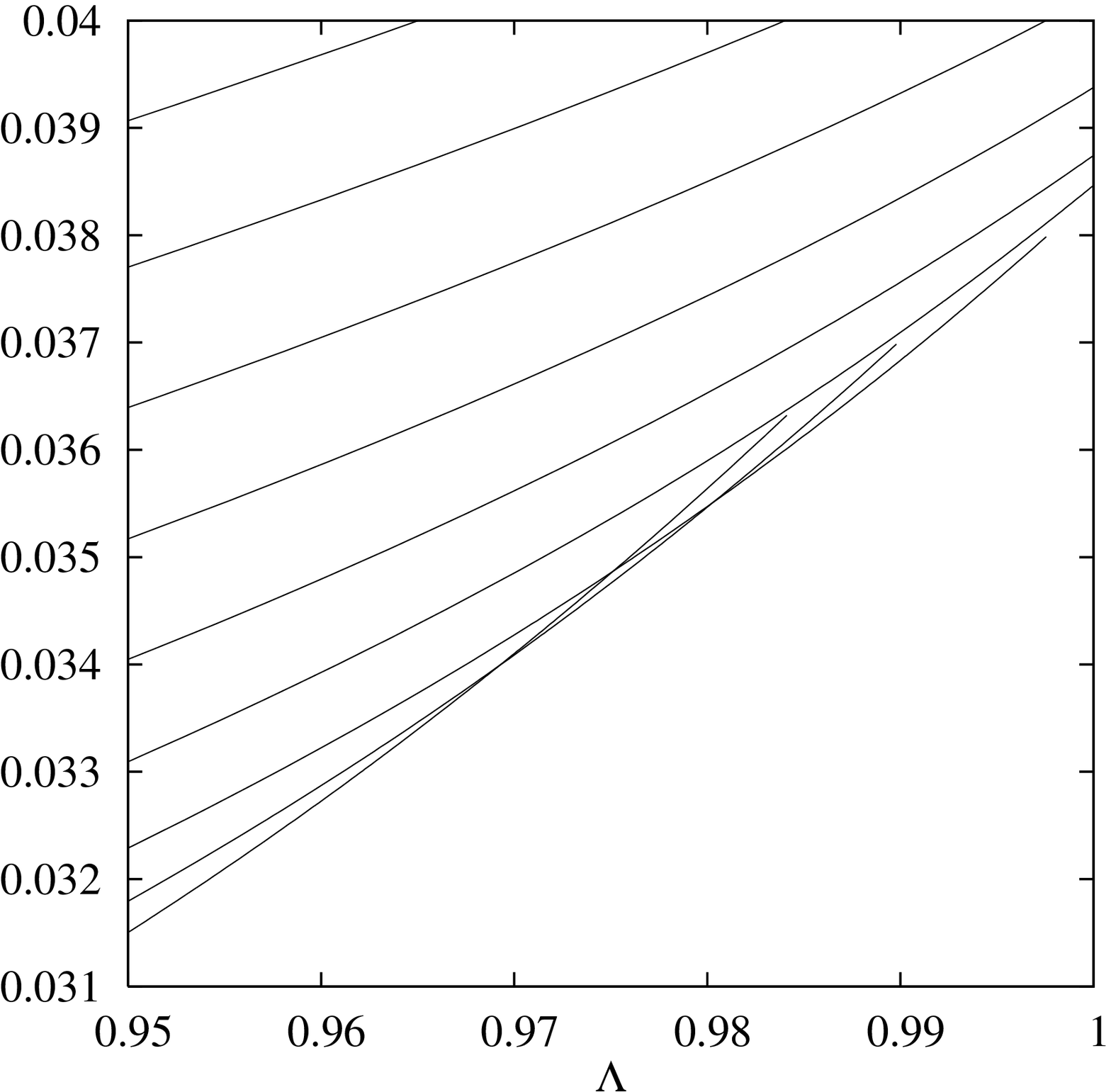}
  \caption{Interaction flows of $\Delta=\Delta_{\rm c}-\chi\Delta_{\rm s.e.}$
    for patch 1 at $U_0=1.5t$, $t^\prime=-0.1t$, $\mu=-1.41t$ (quarter
    filling), $\Delta_{\rm c}$ ranges from $0.05t$ to $0.19t$ and can
    be read off at the intersection of flow line and $\Delta$-axis. }
  \label{fig:Deffiaflows}
\end{figure}

In this section, we employ the interaction flow cutoff
\eqref{eq:iacutoff} with an initial anomalous
self-energy $\Delta_{\rm s.e.,i}=\Delta_{\rm ext}$ and a counterterm 
$\Delta_{\rm c}=\Delta_{\rm ext}$ which we include in the bare propagator.
In figure~\ref{fig:Deffiaflows}, 
we show examples of the flow of the superconducting order parameter
%$\Delta_{\rm eff}:=\Delta_{\rm c}-\chi\Delta$ for various counterterms. 
$\Delta=\Delta_{\rm c}-\chi\Delta_{\rm s.e.}$ for various counterterms. 
We note that if we could treat the flow equations without truncation 
and further approximations, all counterterms should produce the same 
final self-energy and vertices. Indeed, the counterterm-fRG for a 
half-filled charge-density-wave mean-field model exhibited a 
single strong attractor for the flow of the 
self-energy \cite{2006NJPh....8..320G}.
 
For the more general model studied here, the truncation error becomes noticeable, and for certain values of the pairing counterterm, the interaction vertices still diverge. 
The flows are terminated if the maximal effective interaction
exceeds four times the bandwidth, i.e., if the flow leaves 
the weak-coupling regime, or if $\Lambda$ reaches $\Lambda_{\rm f}=1$.
We see in the left diagram that the flows no longer converge onto
a single strong attractor as for the mean-field case
\cite{2006NJPh....8..320G}. Studying the final 5$\%$ of the flow
shown in the right plot of figure~\ref{fig:Deffiaflows}
reveals that there exists a set of small counterterms
for which the flows cross and leave the weak-coupling regime,
and a minimal counterterm
for which the effective interactions remain within the above
limit up to $\Lambda_{\rm f}=1$.
The flows for counterterms whose strength is just beyond
this minimal counterterm terminate closer to each other
than do the flows for larger counterterms.
The minimal counterterm is approximately
twice as strong as the final order parameter value
obtained in the minimal counterterm flow. 
The behavior described above is interpreted as being analogous to the
strong-attractor behavior in the mean-field case. 
The overestimation of the final order parameter value
for larger counterterms is attributed to the overestimation
of the order parameter in the flow due to the neglect of the order parameter's 
energy dependence and momentum dependence perpendicular to the Fermi surface.
Consequently, we choose the minimal order parameter
obtainable by terminating the flow at $\Lambda_{\rm f}$
in the weak-coupling regime as our approximation for the
physical order parameter. 
Another way to argue for the so chosen 
counterterm is that the continuous symmetry 
breaking necessitates a divergent vertex {\em exactly} at the end of 
the flow in order to render the Goldstone mode correctly.
This infinite value is suppressed by our approximation, but
the flow for the chosen counterterm yields
the maximal final value for the Goldstone mode within the above limit.
Finally, note that the order parameter obtained 
by these rules is approximately $20\%$ smaller
than the value obtained with the momentum-shell
method in section \ref{sec:momentumflow}, 
which is enhanced by a finite, albeit small, external field.

\subsubsection{Order parameters}
\begin{figure}[htbp]
  \centering
  \includegraphics[scale=0.5]{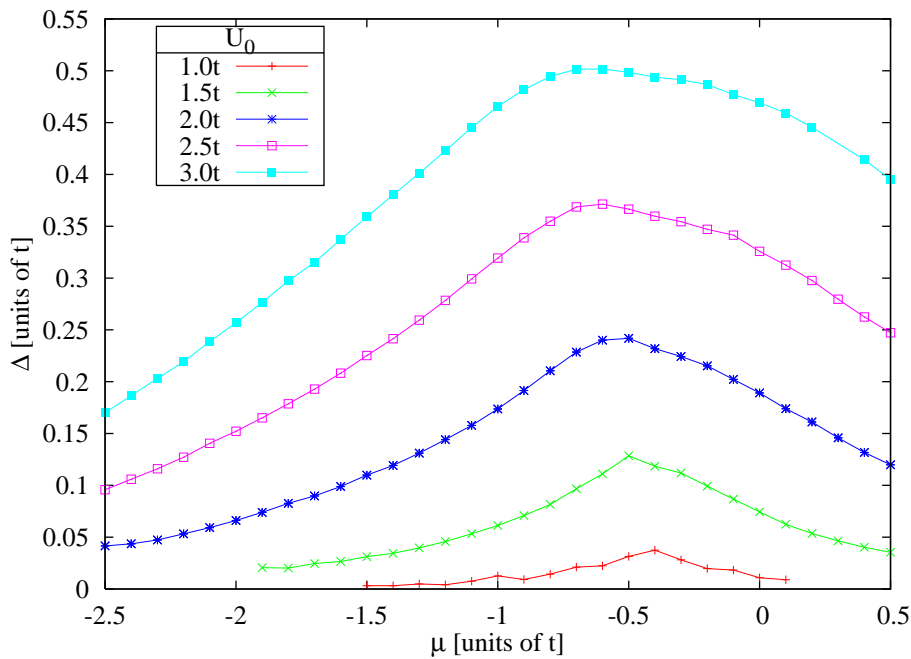}
  \caption{Magnitude of the superconducting order parameter $\Delta$
    as calculated via the interaction flow procedure
    for varying interaction strength $U_0$ and chemical 
    potential $\mu$, $t^\prime=-0.1t$}
  \label{fig:muscans}
\end{figure}

\begin{figure}[htbp]
  \centering
  \includegraphics[scale=.3]{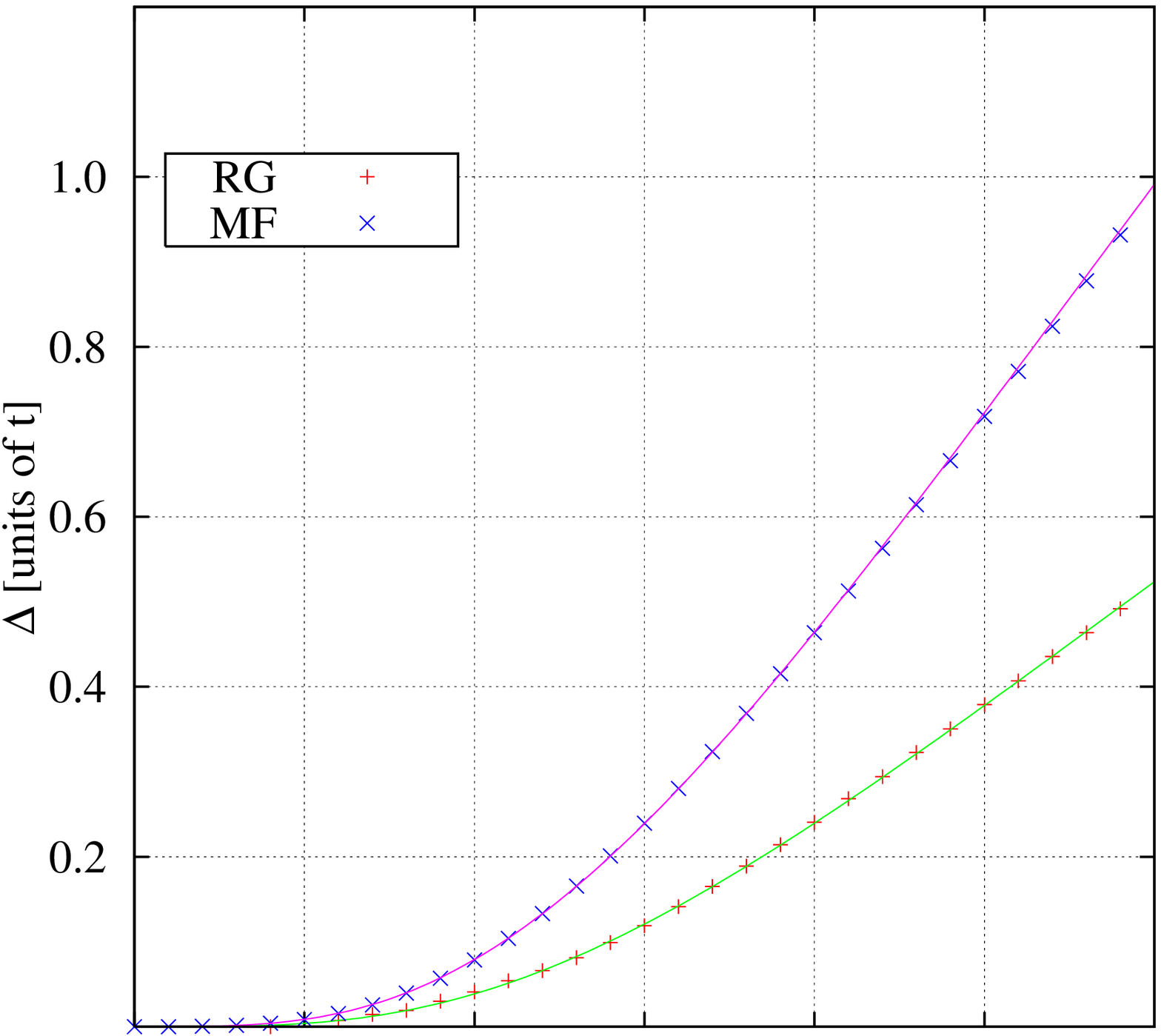}
  \includegraphics[scale=.3]{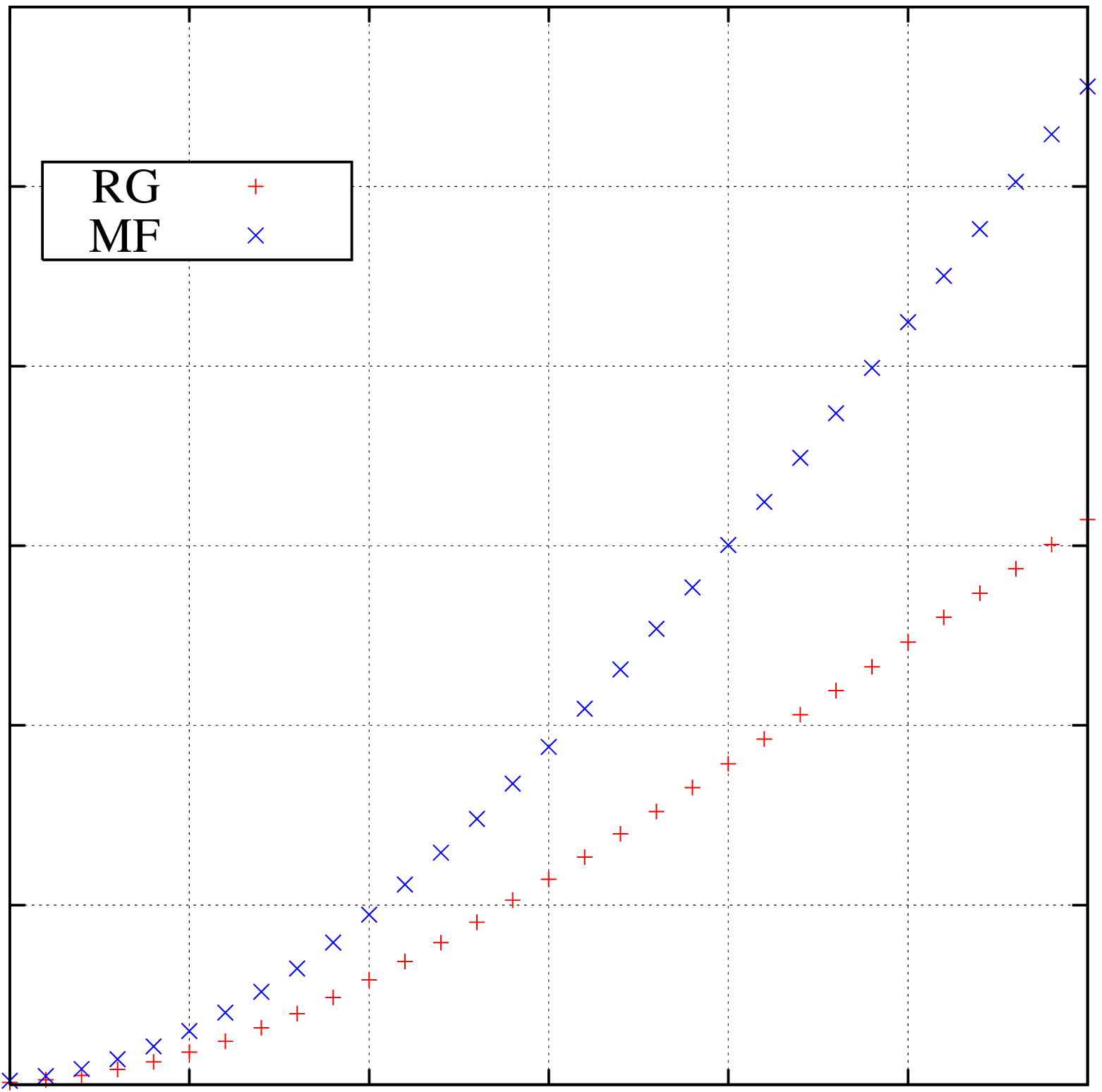}
  \caption{Dependence of the superconducting gap on the effective interaction
    as calculated via the interaction flow procedure, 
    quarter filling: $t^\prime=-0.1t$, $\mu=-1.41t$ (left),
    half filling:$t^\prime=0$, $\mu=0$ (right).
    For the quarter-filled case, fits to $\alpha\exp(-\beta/U_0)$
    are included, where $\alpha$ and $\beta$ are real fit parameters.}
  \label{fig:DvU}
\end{figure}

figure~\ref{fig:muscans}
shows the magnitude of the superconducting order parameter $\Delta$
versus the chemical potential as calculated using the interaction
flow method including a counterterm as described in the 
previous section.
For small bare coupling $U_0$, $\Delta$ is maximal if the 
van Hove points lie on the bare Fermi surface, which is the
case for $\mu=4t^\prime$.
For larger bare coupling, the maximum gradually shifts
to smaller $\mu$.
We compare the results from the fRG 
to results from mean-field theory in figure~\ref{fig:DvU}.
We note that the suppression of the mean-field values by the
fluctuations taken into account by the fRG varies
between a factor 1.6 and 2.0.
This is quite similar to values found by second-order perturbation
theory \cite{PhysRevB.45.13008}.
Table \ref{tab:Julius}
\begin{table}[htbp]
  \centering
  \caption{  \label{tab:Julius}
    Superconducting order parameter $\Delta$ in units of t 
    for $t^\prime = -0.1t$, $\mu=-0.2612t$ (half filling),
    comparing fRG results to results from a combination of
    fRG and MF.
    $^*$: value obtained using linear extrapolation.}
  \begin{indented}
  \item[]\begin{tabular}{|c|c|c|}
    \hline
    U [t] & fRG & fRG and MF \\
    \hline
    1.0 & 0.02 & 0.02 \\
    1.5 & 0.11 & 0.07 \\
    2.0 & 0.22 & 0.15 \\
    2.5 & 0.35 & 0.26 \\
    3.0 & 0.49 & 0.38$^*$ \\
    \hline
  \end{tabular}
\end{indented}
\end{table}
compares our results to results from a combination
of fRG and mean-field theory \cite{juliusdiss}
for finite $t^\prime$. 
While the values are of the same order of magnitude,
the results obtained by employing only the fRG are 
larger by between $20-30\%$.

\section{Conclusions}
\label{sec:conclusions}
We have applied the fermionic one-particle irreducible (1PI)
version of the functional renormalization group (fRG) in the Katanin
truncation to the symmetry-broken
phase of a two-dimensional, non-mean-field model exhibiting 
superconductivity at zero temperature.
The attractive Hubbard model has been chosen as the prototype
system because it is sufficiently
well-studied to provide a good testbed for new many-body methods.
In this case, particle-hole as well as particle-particle loops 
are important for the symmetry-breaking properties of the system.
The fRG takes both into account on an equal footing.

Our analysis has proceeded in three steps.
First, the usual Nambu formalism has been extended
to cover $3+1$ anomalous effective interactions featuring
an odd number of incoming legs,
whose behavior and importance was a focus of our interest.
We have found that for the case of superconductivity,
%the propagator can be denoted as a $2\times2$ matrix in Nambu space.
the effective interaction had to be extended
into a $4\times4$ matrix containing certain redundancies due
to symmetries.
This extension followed from a substitution of the
fields in which the model is originally formulated by Nambu
fields which are chosen specifically to produce an
action of canonical shape while encompassing singlet Cooper
pairs in the quadratic part.
$3+1$ effective interactions were found to arise naturally
within this formalism.

In the second step, a study of the system has been performed
by means of a resummation of a subset of all perturbation theory diagrams.
The strength of the
symmetry breaking was determined by a mean-field gap equation.
A class of diagrams containing bubble chains, ladders, and
all chains comprising both was resummed into a Bethe-Salpeter equation.
In the context of this equation, the impact of the $3+1$ 
effective interactions was studied by comparing effective
interaction values calculated with and without the $3+1$
effective interactions.
This impact was found to be zero for the zero-frequency part of the
phase mode, and to be dependent on the external field 
for non-zero frequency for both phase and amplitude mode.

In the third step, the fRG has been numerically applied
to the attractive Hubbard model.
Two distinct variants of the fRG were employed.
First, a momentum-shell cutoff was used together with an external
field providing a finite initial order parameter.
The Goldstone mode divergence is regularized by the field.
Despite the approximate nature of the flow and
the explosive character of the fRG differential equation, 
flows reaching down to zero scale were found employing external field strengths
two orders of magnitude smaller than the final order parameter values.
{
The relative error of the results for the order parameter due to such
an external field was estimated to be
at most $20\%$, judging from earlier calculations for mean-field models.
}
For even smaller external field values, a regime of diverging
flows was found.
Second, we have calculated interaction flows, balancing an external
field with a counterterm increasing continuously with the flow.
We have noted that vestiges of strong-attractor behavior
found previously for a mean-field model 
%\cite{2006NJPh....8..320G} 
are also present for the attractive Hubbard model flows.
They permit the determination of order parameter values which are
comparable to results from the literature.
While these values proved to be smaller than values calculated
by mean-field theory alone, as expected, 
they are generally slightly larger than
results obtained by combining symmetric-phase Wick-ordered fRG calculations
and mean-field theory.
This deviation is likely due to small differences between
the 1PI and Wick-ordered fRG schemes, since symmetric-phase
1PI calculations already yield larger critical scales than
the corresponding Wick-ordered calculations.
In summary, the 1PI fRG in the Katanin truncation 
was found to yield fairly accurate quantitative
results for weak and moderate interaction strengths (up to $U_0/t=3$).
% {We consider these results as first steps on the way to 
% a fRG scheme that allows the quantitative k-dependent determination of 
% the order parameter of broken-symmetry ground states in interacting 
% fermion systems.  
We reckon that the diverging flows
present for very small external fields
in our calculations can be pushed back by improving the resolution of
the discretization in both frequency and momentum space.
This would furthermore improve the accuracy of the results.
The method may also be employed to study symmetry breaking
in interacting-fermion models for which a stronger momentum
dependence of the order parameter is expected, for example to
study the repulsive Hubbard model.
The method can also be applied to situations with competing order.
The interaction-flow scheme with possible
symmetry-breaking in two different channels has recently been tested in
a one-dimensional model with competing long-range correlations. 
There it
allowed for a correct determination of a quantum critical point between
a charge-gapped phase and a compressible phase with dominant
superconducting correlations \cite{matthias_competing}.
Finally, it will also be interesting to analyze how non-zero temperature 
affects the renormalization of the mean-field-like instability 
encountered in the flows.

%%% Local Variables: 
%%% mode: latex
%%% TeX-master: "main"
%%% End: 

\section*{Acknowledgments}
We thank Andrey Katanin, Jutta Ortloff, Matthias Ossadnik, Julius Reiss, 
Manfred Salmhofer, and Philipp Strack for pleasant and valuable 
discussions, and
Pawel Jakubczyk for providing very useful comments on the
manuscript.\\[.5cm]

\bibliographystyle{unsrt}
\bibliography{main}

\end{document}